# A Behavioral Scorecard Model Using Survival Analysis

Cheng Lee
Independent Analyst
Lee.abcde@gmail.com

Hsi Lee
Department of Management Science
National Yang Ming Chiao Tung University
Hsinchu, Taiwan
doris20030511@gmail.com

## 1. Introduction

Credit risk assessment is the cornerstone of sound lending, allowing institutions to predict defaults and make informed financial choices. Traditionally, logistic regression has been the industry standard for scorecard development, prized for its simplicity and clear-cut binary predictions. However, survival analysis adds a vital second dimension: timing. By integrating these two methodologies, lenders can move beyond asking if a borrower will default to asking when. This hybrid approach creates a more dynamic risk profile, empowering institutions to manage credit proactively rather than reactively.

This article details the development of a monthly hazard rate model that integrates logistic regression with survival analysis techniques to account for time-varying risk factors. Utilizing an augmented longitudinal dataset, the study outlines the end-to-end process of data preparation, model construction, and performance evaluation. By converting discrete-time hazard rates into cumulative default probabilities, the framework culminates in the design of a behavioral scorecard refined through offset adjustment.

## 2. Literature Review

In credit risk modeling, the objective is often to predict the "time-to-default." While the Cox Proportional Hazards model is the "gold standard" in medical biostatistics, the banking industry has historically gravitated toward Logistic Regression for the following advantages:

- Handling Tied Events: It manages tied events more effectively than many continuous-time models (Allison, 2010).
- Computational Efficiency: Parameter estimates are obtained by maximizing the log-likelihood function, which is computationally convenient (Allison, 2010).
- Time-Varying Predictors: By using multiple rows for each loan ID to reflect changing values, it incorporates time-varying predictors into person-level datasets (Singer & Willett, 2003).
- Interpretability: Its outcome is interpreted directly as a hazard rate (probability), whereas the Cox model's primary outcome is a hazard ratio.

Since the banking industry already relies heavily on logistic regression for propensity models, such as mail campaigns, origination scorecards, and cross-selling, this familiar framework is easily applied to discrete-time survival analysis on panel data.

Despite its prevalence in financial institutions, logistic regression has significant limitations. First, the standard data setup relies on discrete monthly, quarterly, or annual intervals, which causes the exact timing of an event within those periods to be lost (Sloma et al., 2021). Second, in contexts like default or fraud detection where events are extremely rare, the model often encounters "separation." This phenomenon leads to a failure of convergence because maximum likelihood estimates may not exist when a predictor perfectly distinguishes the outcomes (Allison, 2012). Third, integrating macroeconomic variables as predictors is challenging due to the model's inherent reliance on historical continuity. These frameworks operate under the rigid assumption that future economic trajectories will mirror past patterns. This dependency becomes a significant liability during periods of unprecedented volatility or structural shifts because the model lacks the flexibility to account for deviations from established norms. Consequently, predictive accuracy diminishes when the economic environment evolves beyond the specific parameters defined by its training data.

## 3. Data

The dataset for this study was sourced from Freddie Mac (www.freddiemac.com) and comprises mortgage loans originated between 2018 and 2021, inclusive of their longitudinal performance histories. This analysis utilizes Data Release 42, with a performance evaluation cutoff date of September 2024. To ensure a homogeneous sample for demonstration purposes, the scope was restricted to 30-year, fixed-rate, single-family primary residence loans.

## 4. Methodology

### 4.1. Data Preparation

Conventional scorecard models typically employ a static performance window of two or three years, with loan status recorded exclusively at the conclusion of this period. Under this framework, each loan in the modeling dataset is represented by a single observation and classified as either "good" or "bad" based on origination-level predictors.

In contrast, this study adopts a survival analysis framework characterized by Type I censoring, as delineated by Lee and Wang (2003), wherein loans are monitored and censored at the 36th month. Loan status is defined as "bad" (default) if the account reaches a delinquency state of 60 days or more, coded as 1, while "good" status is coded as 0. The observation period spans from February 2018 to September 2024, with the longitudinal distribution of loan statuses detailed in Appendix I. As described in Section 3, the final origination date is December 2021. Consequently, the total number of observations declines after this point due to payoffs or defaults.

After removing loans with non-consecutive performance months, the resulting counts are summarized in Table 1.

| Year | # Loans | # Goods | # Bads | # Observations |
|---|---|---|---|---|
| 2018 | 357,132 | 338,861 | 18,271 | 9,449,177 |
| 2019 | 403,324 | 380,363 | 22,961 | 9,961,782 |
| 2020 | 501,132 | 487,459 | 13,673 | 16,081,716 |
| 2021 | 613,717 | 597,730 | 15,987 | 20,926,156 |

Table 1: Numbers of Observations of Overall Data

Following a stratified sampling approach based on loan performance (good or bad), the dataset was partitioned into a training set (70%) and a testing set (30%). This stratification ensures that the relative proportions of "good" and "bad" loans are preserved across both subsets. The resulting distribution of loan statuses within each partition is detailed in Table 2.

| Data | Year | # Loans | # Goods | # Bads | # Observations |
|---|---|---|---|---|---|
| Train | 2018 | 249,993 | 237,203 | 12,790 | 6,611,844 |
| | 2019 | 282,328 | 266,255 | 16,073 | 6,974,040 |
| | 2020 | 350,794 | 341,222 | 9,572 | 11,255,065 |
| | 2021 | 429,602 | 418,411 | 11,191 | 14,649,178 |
| Test | 2018 | 107,139 | 101,658 | 5,481 | 2,837,333 |
| | 2019 | 120,996 | 114,108 | 6,888 | 2,987,742 |
| | 2020 | 150,338 | 146,237 | 4,101 | 4,826,651 |
| | 2021 | 184,115 | 179,319 | 4,796 | 6,276,978 |

Table 2: Numbers of Observations of Train and Test data.

Using the training data shown in Table 2, the in-sample set consists of loans originated between February 2018 and June 2021. Similarly, a holdout sample was created from the test data covering the same period. The observation counts for both sets are detailed in Table 3.

| Sample | # Bads | # Goods | # Total |
|---|---|---|---|
| In-Sample | 41,786 | 29,818,815 | 29,860,601 |
| Holdout | 17,919 | 12,77,5321 | 12,793,240 |

Table 3: Number of Observations of Data

## 4.2. Data Augmentation

The construction of exploded panel data via snapshots creates a structured framework designed to replicate the exact conditions of real-time forecasting. Each historical snapshot functions as a fixed observation point that represents a theoretical "today" within the training set. Data points occurring after this reference date are reorganized so that the independent variable, loan-age, reflects the specific number of months elapsed from that snapshot. This transformation ensures the model views time as a relative horizon rather than a static calendar date. Consequently, when a forecast is required from a future month, that specific month serves as a fresh snapshot and the model projects outcomes based on the incremental loan-age progression from that new starting point.

To augment the original dataset, this study employs the exploded panel data technique (Lynch et al., 2023; Lee, 2023). This approach was originally established by Van Houwelingen and Putter (2011), and its various adaptations have since become standard practice within the banking industry for credit risk assessment.

Consider a loan with a five-month performance history. The expansion process begins by constructing five distinct panels: the first encompasses months one through five, while each subsequent panel initiates from the following month and terminates at month five. This iterative process results in panels of decreasing length, specifically five, four, three, two, and one observation, respectively. When these panels are stacked, they form a fully exploded longitudinal dataset. Within this framework, the initial month of each panel is defined as the snapshot (or landmark) month, and a loan-age covariate is introduced, initialized at zero for each panel, to capture the elapsed duration of the loan. In this specific instance, a record with five original observations generates a total of 15 observations. More generally, a loan characterized by $n$ performance months produces $n(n + 1) / 2$ observations in the fully exploded panel dataset. A comprehensive discussion of these constructions and their associated variations is provided by Lee (2023).

As indicated in Table 3, the original in-sample dataset comprises approximately 29.8 million observations. Constructing a fully exploded panel from a dataset of this magnitude would yield an impractical volume of data, leading to significant computational inefficiencies during model estimation. To mitigate this burden, a progressive weighting scheme is employed to generate a backward weighted sample, following the methodology proposed by Lee (2023). Utilizing the distributions of good and bad loans detailed in Appendix II, the specific algorithm for deriving this backward sample from the primary training data is articulated in Appendix III. The bin sizes for "goods" and "bads" and the associated weights are currently arbitrary; therefore, investigating an optimal methodology for binning and weighting is suggested for future research.

Table 4 provides the observation counts for the original in-sample dataset (comprising originations through June 2021), the fully exploded panel dataset, and the backward weighted sample. This comparison demonstrates that conducting model estimation on a fully exploded dataset of 504 million observations is computationally impractical. Conversely, utilizing the backward weighted sample of 30.6 million observations is significantly more manageable. This approach remains robust because the sample is constructed such that the sum of the weights approximates the total volume of the fully exploded dataset, thereby preserving the statistical power of the larger population.

| Data | Number of Observations | Sum of Weights |
|---|---|---|
| Original In-Sample | 29,860,601 | Not available |
| Full Exploded Panel from In-Sample | 504,557,108 | Not available |
| Backward Weighted Sample of Full Exploded Panel | 30,613,382 | 500,489,652 |

Table 4: Number of Observations of Data

## 4.3. Static and Dynamic Independent Variables

The model incorporates standard loan attributes measured at origination, including the FICO score, interest rate (orig_int_rt), monthly payment, combined loan-to-value (CLTV) ratio, and debt-to-income (DTI) ratio. As these attributes remain constant throughout the duration of the loan performance history, they are classified as static or time-invariant variables.

To account for the influence of the broader economic environment, macroeconomic variables are incorporated as exogenous predictors within the model. These data are sourced from the Federal Reserve Bank (FRB) via the Federal Reserve Economic Data (FRED) repository. A comprehensive list of the macroeconomic variables evaluated in this analysis, along with their respective definitions, is presented in Table 5.

| Variable Name in FRB | Description |
|---|---|
| TDSP | Household Debt Service Payments as a Percent of Disposable Personal Income |
| CDSP | Consumer Debt Service Payments as a Percent of Disposable Personal Income |
| RCMFLBACTDPDPCT30P | 30 or More Days Past Due: account percentage |
| RCMFLBBALDPDPCT30P | 30 or More Days Past Due: balance percentage |
| RCMFLBACTDPDPCT60P | 60 or More Days Past Due: account percentage |
| RCMFLBBALDPDPCT60P | 60 or More Days Past Due: balance percentage |
| RCMFLBACTDPDPCT90P | 90 or More Days Past Due: account percentage |
| RCMFLBBALDPDPCT90P | 90 or More Days Past Due: balance percentage |
| QBPLNTLN3089DU | Loan Performance: Total Loans and Leases: 30-89 Days Past Due |

| QBPLNTLN90DU | Loan Performance: Total Loans and Leases: 90 Days or More Past Due |
|---|---|
| CORCCACBS | Charge-Off Rate on Credit Card Loans |
| CORCACBS | Charge-Off Rate on Consumer Loans |
| CORALACBS | Charge-Off Rate on All Loans |
| CORSFRMACBS | Charge-Off Rate on Single Family Residential Mortgage |
| CORSREACBS | Charge-Off Rate on Loans Secured by Real Estate |
| COROCLACBS | Charge-Off Rate on Other Consumer Loans |
| QBPLNTLNNTCGOFFR | Loan Performance: Total Loans and Leases: Net Charge-Off Rate |
| NCOALLCCACB | Asset Quality Measures, Net Charge-Offs on All Loans and Leases, To Consumers, Credit Cards |
| NCOALLACB | Asset Quality Measures, Net Charge-Offs on All Loans and Leases |
| CPIAUCNS | Consumer Price Index for All Urban Consumers |
| CSUSHPINSA | S&P CoreLogic Case-Shiller U.S. National Home Price Index |
| MORTGAGE30US | For 30-Year Fixed Rate Mortgage Average |
| GDP | Gross Domestic Product |
| UNRATENSA | Unemployment Rate |
| IR10000 | Import Price Index: Crude Oil |
| POILBREUSDM | Global price of Brent Crude |
| POILWTIUSDM | Global price of WTI Crude |

Table 5: Macroeconomic Variables Considered

In the context of mortgage modeling, the Spread at Origination (SATO) is widely recognized as a significant risk factor. It is formally defined as the differential between the loan's specific origination interest rate and the prevailing market average for 30-year fixed-rate mortgages, as represented by the Federal Reserve Bank series MORTGAGE30US.

The selection of candidate loan-level attributes and macroeconomic variables is informed by the foundational frameworks established by Lee and Lee (2010) and Bellini (2019).

In addition to loan-level attributes and macroeconomic variables, the model incorporates three duration-related variables: months on book (MOB), remaining term (mths_remng), and loan age (loan_age). MOB quantifies the total duration for which a loan has been active since origination, whereas mths_remng indicates the number of months until maturity. Loan age is defined as a dynamic covariate that resets to zero for each snapshot month and represents the elapsed time within a specific subpanel. As delineated in Section 4.2, a snapshot dataset constitutes a subpanel within the larger exploded panel structure. Consequently, loan age initializes at zero in each snapshot; MOB initializes at zero at the time of origination; and mths_remng is a monotonically decreasing predictor with a starting value of 360.

## 4.4. Variable Transformations and Section

To evaluate the relationship between the dependent variable and each predictor, bivariate plots illustrating the log-odds of default against individual loan attributes are presented in Appendix IV. Based on these visualizations, original unpaid balances (orig_upb) is excluded from the model due to a non-monotonic and counterintuitive relationship with the log-odds. With the exceptions of FICO, DTI, interest rate, and SATO, several attributes exhibit nonlinear associations with the log-odds. Consequently, spline transformations are applied to loan age (loan_age), months on book (MOB), remaining term (mths_remng), and CLTV.

These splines are constructed using multivariate adaptive regression splines (MARS), following the methodology described by Hastie et al. (2017). The MARS framework captures nonlinear effects through piecewise linear basis functions of the form $(x - c)_+$ or $(c - x)_+$ where $x$ represents the covariate, c denotes a specific knot location, and the subscript "+" indicates the positive part of the function. Consistent with the bivariate analysis in Appendix IV, the knot locations identified for loan age, MOB, CLTV, and monthly payment are reported in Table 6. No spline transformation is applied to the interest rate (orig_int_rt), as its relationship with the log-odds remains approximately linear.

| Variable | Knot Locations |
|---|---|
| Loan_Age | 1, 22 |
| MOB | 3, 23, 27 |
| Mths_remng | 334, 358 |
| CLTV | 82 |

Table 6: Knot Locations of Variables

Regarding the macroeconomic variables, several transformations are analyzed, including year-over-year differences, quarter-over-quarter differences, and their respective percentage changes. Following an examination of the relationship between each transformation and the log-odds of default, two specific indicators are selected: the quarter-over-quarter percentage change of the 90-day delinquency rate (RCMFLBACTDPDPCT90P) and the one-month lag of the unemployment rate (UNRATENSA). These selections are informed by the bivariate analysis provided in Appendix IV, which illustrates the predictive strength of these specific forms.

## 4.5. Interaction

Several interaction effects are evaluated within the modeling framework. To illustrate this approach, the variable fico_orig_upb is constructed to capture the interaction between the FICO score and original unpaid balances (Orig_UpB). An empirical interaction effect becomes evident when analyzing average default rates segmented by both FICO and Orig_UpB. As demonstrated in the figures below, the relationship between payment size and default risk is conditioned upon the credit score of the borrower. Appendix V provides a comprehensive description of the interaction construction and the underlying data partitions.

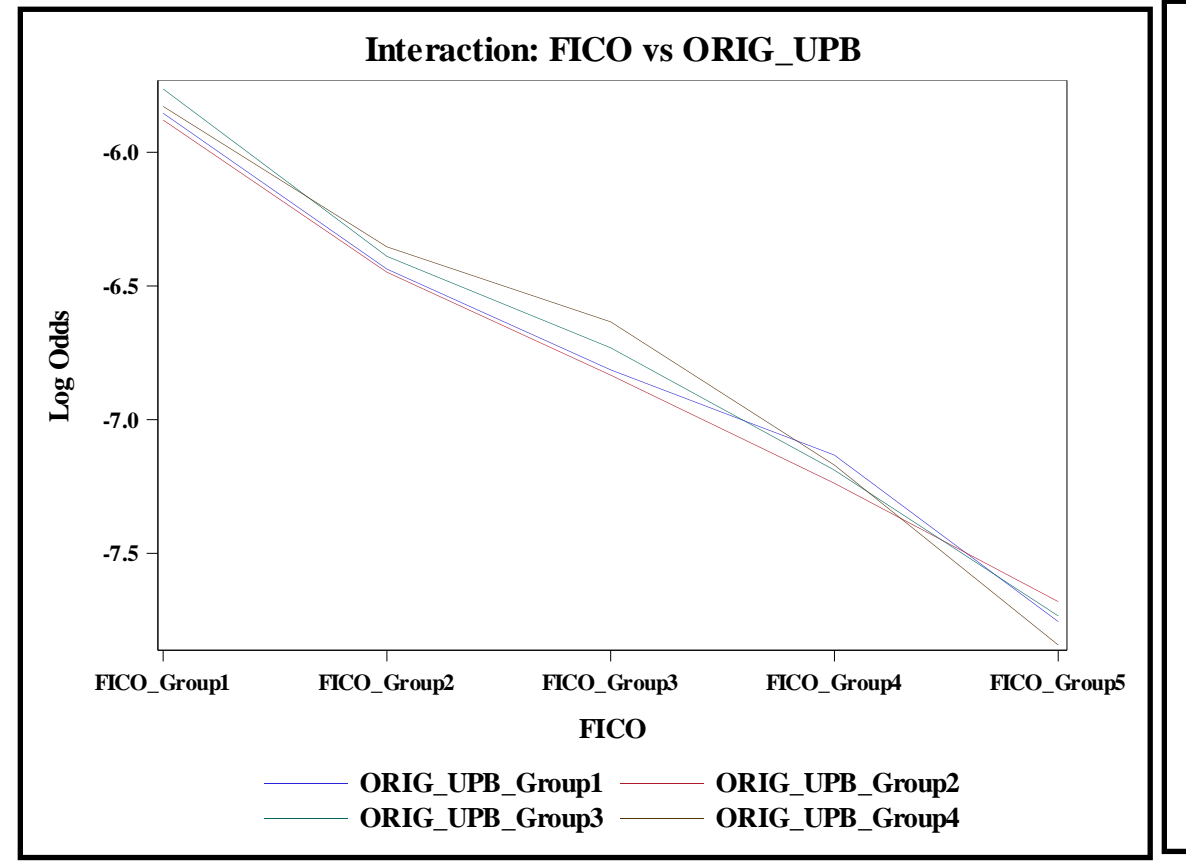

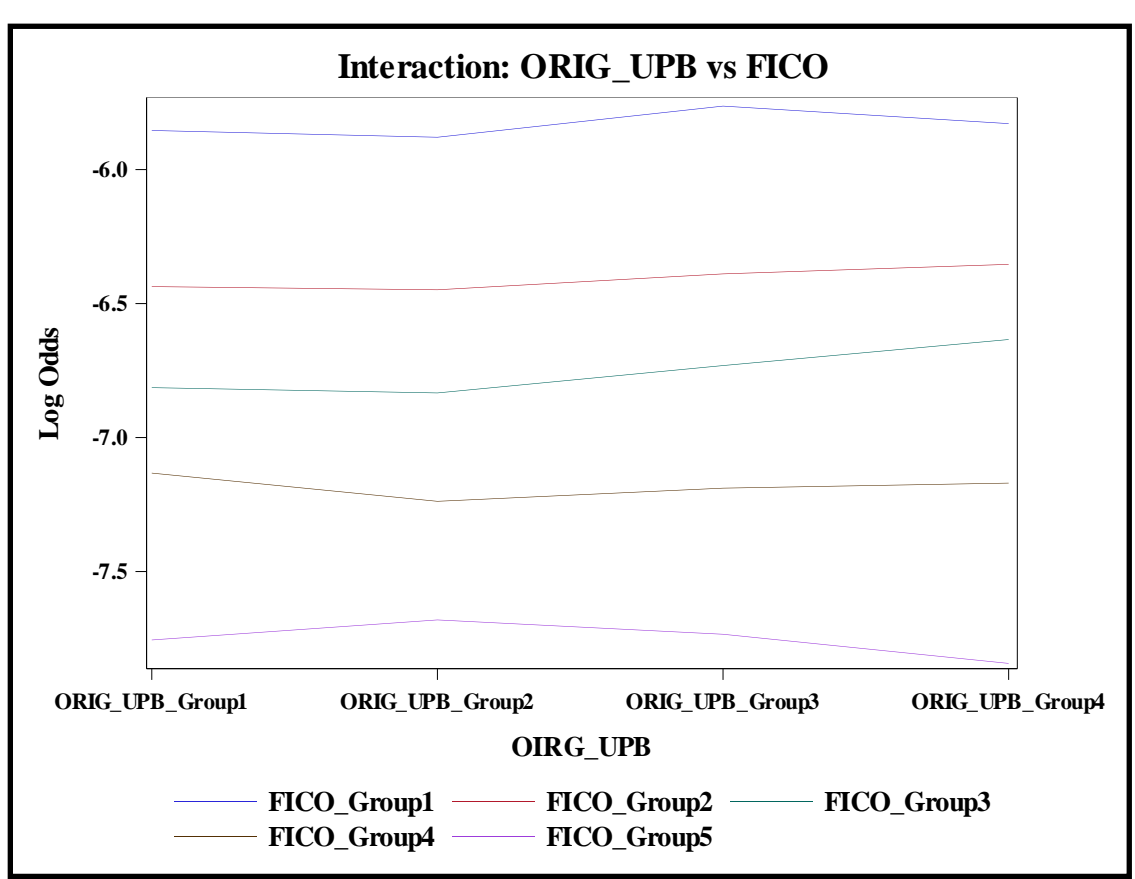

## 4.6. Seasonality

To account for seasonality, the model incorporates monthly indicator variables. These binary predictors are constructed for each month of the year, with December (month 12) serving as the reference category.

## 5. Final Model

### 5.1. Parameter Estimates of The Final Model

Logistic regression is employed to estimate the monthly default rate, which serves as the hazard rate within the survival analysis framework. In this context, the hazard rate represents the conditional probability of default in a specific month, given that the loan has remained active and non-defaulted through the preceding month. The methodology of integrating Multivariate Adaptive Regression Splines (MARS) or Classification and Regression Trees (CART), alongside additional predictors, into a logistic regression framework follows the approaches established by Kuhnert et al. (2000) and Taniar (2008). The parameter estimates for the statistically significant independent variables are reported in Table 7.

| Parameter | Estimate | Standard Error | Wald Chi-Square | Pr > ChiSq |
|---|---|---|---|---|
| Intercept | -2.3877 | 0.0589 | 1641.791 | <.0001 |
| loan_age_Basis (1 – loan_age)$_{+}$ | -14.2841 | 6.1804 | 5.3416 | 0.0208 |
| mob_Basis1 (23 – MOB)$_{+}$ | 0.0144 | 0.00136 | 111.769 | <.0001 |
| mob_Basis2 (MOB – 27)$_{+}$ | -0.0202 | 0.000726 | 771.2851 | <.0001 |
| mths_remng_Basis1 (mths_remng – 334)$_{+}$ | -0.0234 | 0.00111 | 444.6272 | <.0001 |
| mths_remng_Basis2 (mths_remng – 358)$_{+}$ | -3.869 | 0.1825 | 449.5349 | <.0001 |
| cltv_Basis1 (CLTV – 82)$_{+}$ | 0.0254 | 0.000226 | 12712.39 | <.0001 |
| cltv_Basis2 (82 – CLTV)$_{+}$ | -0.00684 | 0.000213 | 1031.882 | <.0001 |
| FICO | -0.00988 | 0.000044 | 51594.68 | <.0001 |
| DTI | 0.0432 | 0.00016 | 72798.1 | <.0001 |
| SATO | 0.1212 | 0.00611 | 393.9696 | <.0001 |
| FICO_ORIG_UPB | -0.00215 | 0.000032 | 4529.852 | <.0001 |
| Orig_Int_Rt | 0.1311 | 0.00517 | 642.4949 | <.0001 |
| UNRATENSA_lag_1month | 0.264 | 0.00103 | 66156.18 | <.0001 |
| RCMFLBACTDPDPCT90P Quarter to Quarter Percentage Change | 0.7794 | 0.0127 | 3737.37 | <.0001 |
| GDP Year to Year Percentage Change | -0.1526 | 0.0606 | 6.3388 | 0.0118 |
| CSUSHPINSA_lag_3month | 0.00428 | 0.000134 | 1026.219 | <.0001 |
| month2 | -0.2474 | 0.0064 | 1494.801 | <.0001 |
| month3 | -0.4652 | 0.007 | 4419.49 | <.0001 |
| month4 | -0.4111 | 0.0076 | 2924.681 | <.0001 |
| month5 | -0.3359 | 0.00671 | 2509.002 | <.0001 |
| month6 | -0.4236 | 0.00662 | 4098.379 | <.0001 |
| month7 | -0.4936 | 0.00613 | 6488.832 | <.0001 |
| month8 | -0.5891 | 0.00629 | 8769.198 | <.0001 |
| month9 | -0.341 | 0.00609 | 3131.192 | <.0001 |
| month10 | -0.1731 | 0.00623 | 773.2177 | <.0001 |
| month11 | 0.014 | 0.00603 | 5.363 | 0.0206 |

Table 7: Parameter Estimates of The Final Model

Based on the estimated coefficients, several predictors exhibit a positive association with the default rate, including the debt-to-income (DTI) ratio, the combined loan-to-value (CLTV) ratio (with the cltv_Basis2 transformation functioning as a monotonically decreasing predictor), SATO, and the origination interest rate. Furthermore, macroeconomic factors such as the quarter-over-quarter percentage change in 90-day delinquency rates, the one-month lag of the unemployment rate, and the three-month lag of the Home Price Index (HPI) demonstrate a positive relationship with default risk, as does the November seasonal indicator.

Conversely, variables that demonstrate a negative relationship with the default rate include the FICO score, the interaction effect between FICO and original unpaid balances (Orig_UpB), and the year-over-year percentage change in GDP. Analysis of the seasonal dummy variables suggests that default rates are significantly higher during November and December relative to the baseline and other months in the calendar year.

## 5.2. Parameter Estimates of GEE

As Lynch et al. (2023) observe, multiple landmark datasets are frequently generated from the same loan. Because observations derived from a single loan may be correlated, Generalized Estimating Equations (GEE) were utilized to fit the model using specified working correlation matrices. During the initial fitting process, the variable loan_age_basis caused data separation and resulted in a non-positive definite Hessian matrix. Although Allison (2012) generally advises against removing influential variables, this specific spline variable was excluded from the GEE application because it lacked predictive power and produced the second-largest *p*-value in the logistic regression.

An alternative approach to mitigate data separation without discarding predictors involves adding additional records to the dataset as proposed by White et al. (2010). The number of observations to be added is determined by the formula $2pk$, where $p$ represents the number of predictors and $k$ represents the number of levels of the target variable. For each target level, two records are created using either $\overline{x_i} + s_i$ or $\overline{x_i} - s_i$, where $\overline{x_i}$ and $s_i$ are the mean and standard deviation of $\overline{x_i}$ while all other predictors are held at their mean values. Each added record is then assigned a weight equal to $(p+1)/2pk$. In this model, $p$ is 26 and $k$ is 2, resulting in 104 records added to each GEE cluster identified by ID and snapshot date. Given that the maximum number of snapshot dates is 37, the final total of added records is 3,848.

Regarding the GEE structure, Agresti (2013) recommends using an exchangeable working correlation matrix when no dramatic correlations are expected. He further suggests that if empirical and model-based standard errors remain similar across all working correlation structures, the correlations among observations are likely modest.

Appendix VI provides the parameter estimates, empirical standard errors, model-based standard errors, and *p*-values derived from fitting the GEE model on both the data excluding loan_age_basis and the data supplemented with additional records. These results encompass various working correlation matrices including independent, exchangeable, and AR(1) structures. Because the empirical and model-based standard errors are similar across these configurations, the correlations among observations within the same loan are considered modest. Consequently, employing a standard logistic regression is appropriate as the additional complexity of the GEE approach is unnecessary.

## 5.3. Backtest

### 5.3.1. In-Sample

Backtest plots for the in-sample and holdout sample are illustrated by date in the figures below. The mean absolute error (MAE) and root mean square error (RMSE) are both negligible, which suggests a high degree of model fit. However, the mean absolute percentage error (MAPE) appears relatively elevated.

This discrepancy arises because MAPE is highly sensitive to observations with near-zero or zero actual values, as the calculation involves division by the actual observed rate.

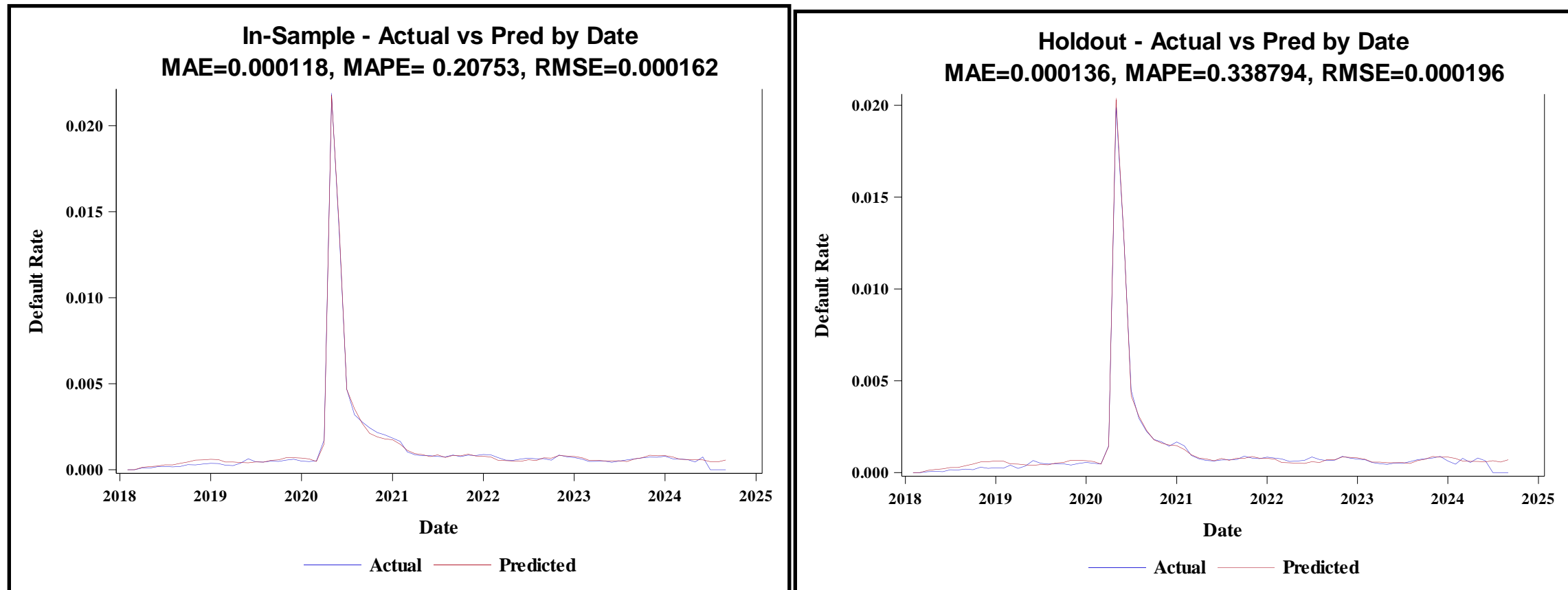

Visualizations of selected snapshots are provided in Appendix VII, illustrating that the predicted values closely align with observed outcomes. A notable spike in June 2020 is attributable to the economic disruptions associated with the COVID-19 pandemic. Because this volatility is effectively captured by the macroeconomic variables already incorporated into the model, the inclusion of a separate COVID-19 indicator is unnecessary.

Furthermore, when snapshots ranging from February 2018 to September 2021 are aggregated, the resulting analysis consistently demonstrates a high degree of alignment between predicted and actual values, as illustrated in the following figure.

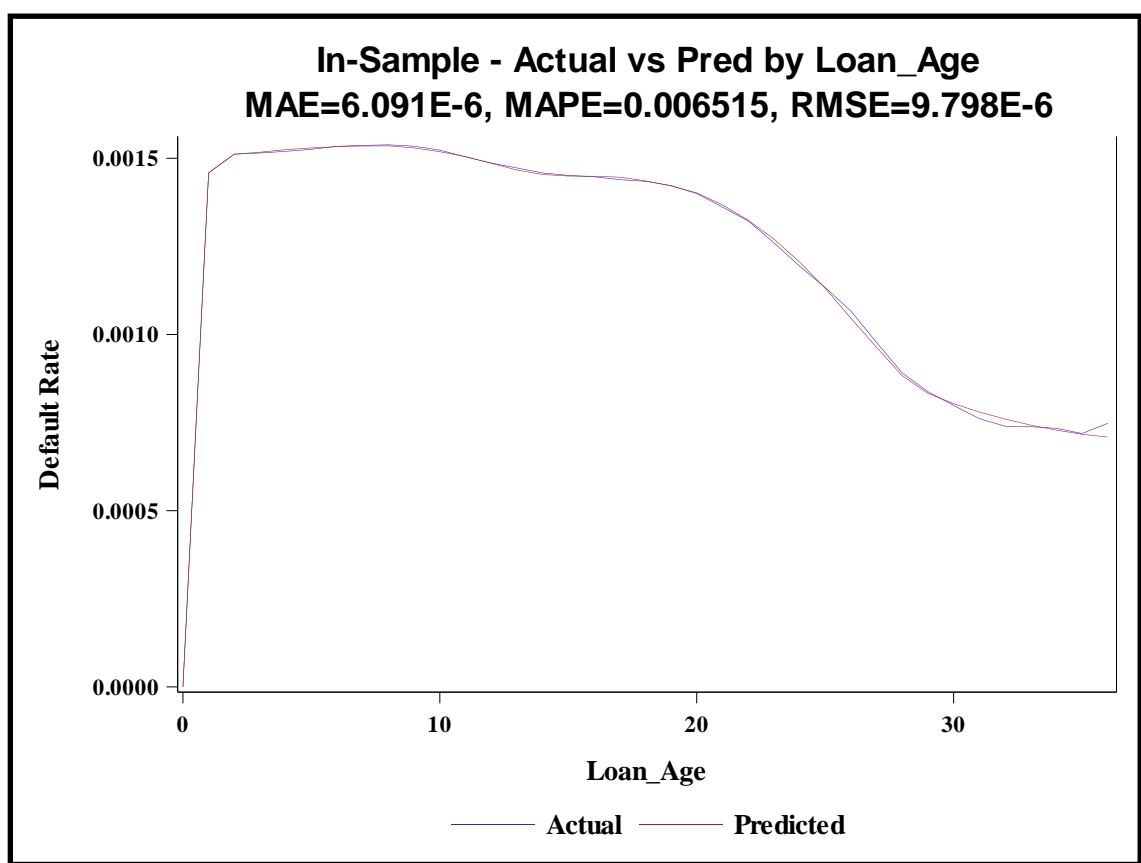

### 5.3.2. Out-of-Time Sample

Loans originated in July 2021 are designated as the out-of-time (OOT) validation sample. The following figure indicates that the predicted hazard rate increases to approximately 0.0007 within the initial three months, exhibits fluctuations between months 4 and 24, and attains a peak of approximately 0.0008 at month 27. The model appears to underpredict default rates during the 2022 and 2023 calendar years. These discrepancies are potentially attributable to shifts in the macroeconomic environment during 2022 and 2023 that diverged from the historical patterns captured within the model development dataset.

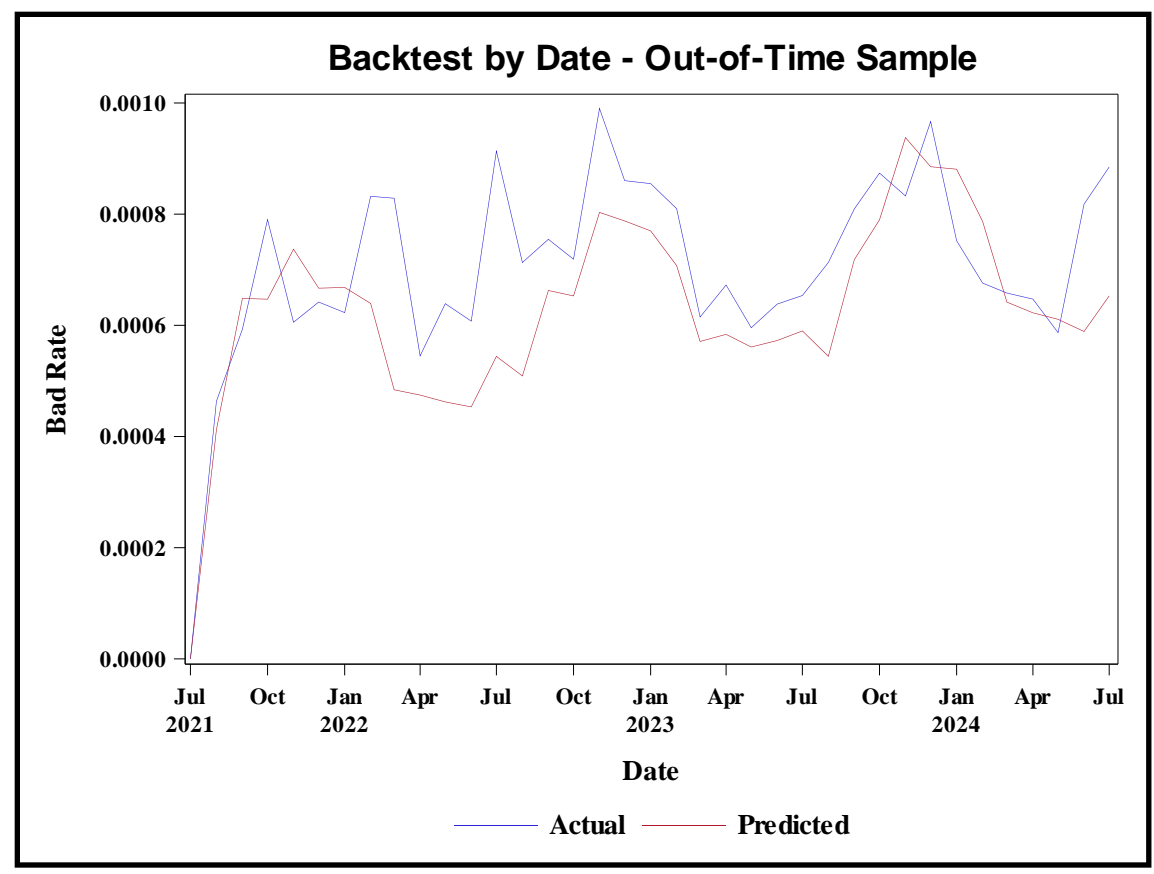

## 6. Model Implementation

### 6.1. Converting Monthly Hazard Rate to Default Probability

Converting a hazard rate to a probability is a fundamental step in survival analysis because it allows researchers to quantify the likelihood of an event occurring within a specific timeframe rather than just at an instantaneous moment. While the hazard rate represents the instantaneous risk or the velocity at which events are occurring at time *t*, it is not a probability itself and can actually exceed 1. This conversion is essential for financial modeling because it translates abstract rates of change into a bounded and intuitive metric ranging from 0 to 1 that represents the actual risk of a default, failure, or event happening over a defined period.

In this specific framework, the model is designed to predict monthly hazard rates over a horizon of up to 36 months. To support comprehensive credit risk assessment, these hazard rates are transformed into default probabilities, which subsequently facilitate the assignment of credit scores to borrowers. The precise methodology for this conversion is outlined in the following steps.

1. Loans with 36 months of performance history, or another specified observation period, are selected from all snapshots. It should be noted that certain loans exhibit fewer than 36 months of history because they were either lost to follow-up or defaulted prior to censoring.
2. By selecting snapshots with a 36-month performance history, or another specified observation period, from the in-sample data, the hazard rates are converted into survival probabilities. This conversion is based on the following formula from Lawless (2003):

$$S(t) = \prod_{j:t_{j<t}} [1 - h(t_j)]$$

The monthly default probability is subsequently obtained by multiplying the hazard rate by the survival probability, as established by the following equation from Lawless (2003):

$$h(t_j) = \frac{f(t_j)}{S(t_j)}$$

### 6.2. Monthly Loss Forecast

The proposed hazard rate model, which is based on exploded panel data, is a standard approach adopted by financial institutions for the Comprehensive Capital Analysis and Review (CCAR). This framework is classified as a probability of default (PD) model because its hazard rates are convertible into default probabilities. When integrated with loss given default (LGD) and exposure at default (EAD) models, it facilitates the estimation of monthly credit losses.

## 6.3. Behavioral Scorecard

The model is constructed within a survival analysis framework in which loans are censored at month 36. This approach enables the development of behavioral scorecards with shorter time horizons, thereby facilitating proactive courtesy calls to mitigate default risk. As defined by Van Gestel and Baesens (2009), a behavioral scorecard is specifically designed to predict the future performance of clients over such condensed observation periods.

### 6.3.1. Performance Window

A scorecard model is typically constructed by monitoring loans over a set performance window and recording the default status at the end of that period. It is therefore appropriate to use the default probability at the final observation point where the terminal status is documented. Each loan is represented by a single record within the scorecard sample. Because the performance window is fixed in a standard scorecard model, a new model must be built whenever a different performance window is required.

The survival analysis framework offers an alternative where the performance window can be flexibly determined based on specific business needs. For illustrative purposes, the following sections employ a six-month observation window to develop a behavioral scorecard that assesses the probability of default within that timeframe. This specific interval is designed to be adjustable so that the model can be recalibrated according to varying institutional requirements or distinct regulatory standards.

### 6.3.2. Offset Adjustment

Following the construction of the scorecard sample described in the preceding section, the counts of "good" and "bad" loans deviate from the proportions observed in the original model. Rather than developing a new model with the updated dataset, the converted probabilities may be recalibrated by incorporating an offset, a method supported by Siddiqi (2017).

Let Sample 1 denote the exploded panel data and Sample 2 represent the scorecard sample. The offset is subsequently calculated as follows:

$$\frac{Number\ of\ Bads\ in\ Sample\ 2 \times Number\ of\ Goods\ in\ Sample\ 1}{Number\ of\ Goods\ in\ Sample\ 2 \times Number\ of\ Bads\ in\ Sample\ 1}.$$

Generally, Sample 2 represents a subset of Sample 1, which served as the original development dataset for the model.

Consequently, $p_2$, the offset-adjusted probability for sample 2 is $\frac{\exp\left(offset+log\left(\frac{p_1}{1-p_1}\right)\right)}{1+exp\left(offset+log\left(\frac{p_1}{1-p_1}\right)\right)}$, where $p_1$is probability from sample 1.

The calculated offset is 1.7925 from the numbers in Table 8.

| Weight Adjusted | Sample 1 | Sample 2 |
|---|---|---|
| Number of Bads | 669,306 | 236,310 |
| Number of Goods | 499,819,830 | 29,389,265 |

Table 8: Number of bads and goods

Rather than developing a new model specifically for Sample 2, the offset method recalibrates the probabilities estimated from the Sample 1 model to approximate the values that would have been obtained if a model were derived directly from Sample 2.

Comprehensive performance metrics, including the Area Under the Curve (AUC), rank-order tables, and backtest plots for both the in-sample and the out-of-time validation sample, are provided in Appendix VIII. The results demonstrate that the offset-adjusted probabilities align closely with observed outcomes. Specifically, the AUC values for the in-sample and out-of-time datasets are 0.82 and 0.70, respectively. These results are classified as "excellent" and "acceptable" based on the evaluative thresholds established by Hosmer et al. (2013).

### 6.3.3. Cutoff Probability for Making Courtesy Calls

Each client within a specified performance window is assigned a predicted probability of default. Given that contacting all clients for courtesy reminders is often operationally impractical, lenders may employ a cutoff or threshold score to prioritize those with the highest delinquency risk. Establishing a predefined threshold enables lenders to streamline decision-making, maintain evaluative consistency, and systematically reduce default exposure.

To determine the optimal threshold, this study utilizes the Youden Index. This metric represents the maximum vertical distance between the diagonal chance line and the Receiver Operating Characteristic (ROC) curve, and it is frequently cited as an optimal classification threshold (Krzanowski & Hand, 2009). In this analysis, the default probability corresponding to the maximum Youden Index is 0.007835.

### 6.3.4. Scaling Default Probabilities to Credit Scores

Because default probabilities may lack intuitive clarity for stakeholders within a line of business, scaling these probabilities into a score format similar to the FICO system is often more beneficial for operational transparency. A widely adopted scaling methodology is the points to double the odds (PDO) approach. This process requires the selection of a baseline score at which the "good-to-bad" odds ratio is doubled for every specified increase in the scaled score. Following common industry practice, an increase of 20 points is utilized to represent a doubling of the odds (Yhip & Alagheband, 2020).

Based on the scorecard sample, the probability at the Youden Index is 0.007835, which corresponds to a good-to-bad odds ratio of 126.63. This ratio is derived by dividing the probability of a "good" status, calculated as 1 – 0.007835, by the probability of default 0.007835. To establish the scaling formula for converting a given default probability into a standardized score, the following two equations are solved using a baseline score of 600.

$$600 = \alpha + \beta \times Log(126.63)$$

$$620 = \alpha + \beta \times Log(2 \times 126.63)$$

By solving for $\alpha$ and $\beta$ in the two equations, the scaled score is expressed as 460.3099 + 28.8539 × Log(odds). It should be noted that the scaling factor $\beta$ is equal to Log(20), where 20 is the standard points to double the odds. For example, a default probability of 0.15 results in a scaled score of 510, calculated as 460.3099 + 28.8539 × Log((1-0.15)/0.15).

The comprehensive procedures for developing this conversion framework are detailed in Yhip and Alagheband (2020). It should be noted that as the default probability approaches zero, the resulting score increases exponentially, necessitating a predefined maximum cap to maintain the practical utility of the scoring scale.

In this study, the baseline score of 600 is established at the probability corresponding to the Youden Index. Consequently, 600 serves as the primary threshold for identifying candidates for courtesy calls. Individuals with a behavioral score below this value are recommended for inclusion on the intervention list. However, this threshold remains flexible and may be calibrated according to the available staffing resources of the institution.

## 7. Conclusions

Utilizing a survival analysis framework, this study initially develops a hazard rate model via logistic regression with data augmentation to estimate monthly default hazards. This methodology is highly extensible and can be adapted to model diverse outcomes such as loan prepayments or product enrollment propensities. Generally, this type of hazard model is referred to as a Probability of Default (PD) model. The process has been widely adopted by financial institutions under the Comprehensive Capital Analysis and Review (CCAR) regulatory framework. Because a financial institution typically manages several products, each may require its own specific CCAR model. Consequently, an institution often maintains multiple CCAR PD models for the primary purpose of undergoing regulatory stress tests.

Without the need to create entirely new models, this study applies an offset adjustment to convert monthly hazard rates into cumulative default probabilities. This approach allows for the construction of a behavioral scorecard to facilitate strategic decision-making throughout the customer lifecycle, including cross-selling initiatives, default mitigation, and retention management.

## Appendix I: Number of Bads and Goods by Month of Overall Data

| Date | # Bads | # Goods | Date | # Bads | # Goods | Date | # Bads | # Goods |
|---|---|---|---|---|---|---|---|---|
| 2/1/2018 | 0 | 17114 | 5/1/2020 | 14171 | 705736 | 8/1/2022 | 787 | 1087415 |
| 3/1/2018 | 0 | 36885 | 6/1/2020 | 9341 | 703904 | 9/1/2022 | 823 | 1064967 |
| 4/1/2018 | 5 | 66779 | 7/1/2020 | 3254 | 723380 | 10/1/2022 | 719 | 1042772 |
| 5/1/2018 | 8 | 98727 | 8/1/2020 | 2109 | 752552 | 11/1/2022 | 980 | 1022296 |
| 6/1/2018 | 15 | 133874 | 9/1/2020 | 1757 | 782933 | 12/1/2022 | 866 | 1002606 |

| 7/1/2018 | 30 | 171568 | 10/1/2020 | 1554 | 813269 | 1/1/2023 | 807 | 985443 |
|---|---|---|---|---|---|---|---|---|
| 8/1/2018 | 33 | 207644 | 11/1/2020 | 1415 | 842199 | 2/1/2023 | 781 | 967904 |
| 9/1/2018 | 42 | 243722 | 12/1/2020 | 1342 | 867236 | 3/1/2023 | 590 | 954056 |
| 10/1/2018 | 61 | 272573 | 1/1/2021 | 1395 | 893117 | 4/1/2023 | 573 | 937646 |
| 11/1/2018 | 88 | 302786 | 2/1/2021 | 1397 | 906581 | 5/1/2023 | 540 | 915146 |
| 12/1/2018 | 98 | 330187 | 3/1/2021 | 857 | 909246 | 6/1/2023 | 528 | 891899 |
| 1/1/2019 | 112 | 352972 | 4/1/2021 | 719 | 920532 | 7/1/2023 | 613 | 866490 |
| 2/1/2019 | 124 | 372564 | 5/1/2021 | 662 | 934669 | 8/1/2023 | 603 | 828479 |
| 3/1/2019 | 118 | 393684 | 6/1/2021 | 670 | 956254 | 9/1/2023 | 604 | 783621 |
| 4/1/2019 | 94 | 420876 | 7/1/2021 | 634 | 997555 | 10/1/2023 | 668 | 736951 |
| 5/1/2019 | 160 | 451189 | 8/1/2021 | 634 | 1038986 | 11/1/2023 | 629 | 688318 |
| 6/1/2019 | 325 | 486055 | 9/1/2021 | 708 | 1079220 | 12/1/2023 | 642 | 640341 |
| 7/1/2019 | 261 | 522574 | 10/1/2021 | 818 | 1113342 | 1/1/2024 | 541 | 597061 |
| 8/1/2019 | 239 | 556677 | 11/1/2021 | 792 | 1144113 | 2/1/2024 | 451 | 551405 |
| 9/1/2019 | 309 | 588334 | 12/1/2021 | 822 | 1169587 | 3/1/2024 | 392 | 518941 |
| 10/1/2019 | 270 | 609632 | 1/1/2022 | 887 | 1196883 | 4/1/2024 | 354 | 487338 |
| 11/1/2019 | 300 | 630211 | 2/1/2022 | 959 | 1182326 | 5/1/2024 | 293 | 444952 |
| 12/1/2019 | 336 | 650267 | 3/1/2022 | 892 | 1169302 | 6/1/2024 | 348 | 405180 |
| 1/1/2020 | 369 | 672543 | 4/1/2022 | 674 | 1155389 | 7/1/2024 | 347 | 362167 |
| 2/1/2020 | 309 | 686985 | 5/1/2022 | 696 | 1140817 | 8/1/2024 | 317 | 300685 |
| 3/1/2020 | 326 | 701046 | 6/1/2022 | 769 | 1124630 | 9/1/2024 | 264 | 241411 |
| 4/1/2020 | 1036 | 714953 | 7/1/2022 | 836 | 1106270 | | | |

## Appendix II: Number of Bads and Goods by Month of In-Sample

The in-sample origination date ends on June 30, 2021; consequently, record counts are lower for the second quarter of 2021.

| Date | # Bads | # Goods | Date | # Bads | # Goods | Date | # Bads | # Goods |
|---|---|---|---|---|---|---|---|---|
| 2/1/2018 | 0 | 12017 | 5/1/2020 | 9880 | 494103 | 8/1/2022 | 324 | 485864 |
| 3/1/2018 | 0 | 25823 | 6/1/2020 | 6582 | 492782 | 9/1/2022 | 311 | 471255 |
| 4/1/2018 | 4 | 46798 | 7/1/2020 | 2279 | 506586 | 10/1/2022 | 263 | 456642 |
| 5/1/2018 | 6 | 69188 | 8/1/2020 | 1440 | 527012 | 11/1/2022 | 395 | 443343 |
| 6/1/2018 | 13 | 93753 | 9/1/2020 | 1229 | 548163 | 12/1/2022 | 335 | 430307 |
| 7/1/2018 | 22 | 120220 | 10/1/2020 | 1113 | 569325 | 1/1/2023 | 311 | 418962 |
| 8/1/2018 | 24 | 145277 | 11/1/2020 | 991 | 589494 | 2/1/2023 | 263 | 407350 |
| 9/1/2018 | 28 | 170640 | 12/1/2020 | 967 | 607069 | 3/1/2023 | 200 | 398239 |
| 10/1/2018 | 48 | 190714 | 1/1/2021 | 947 | 625207 | 4/1/2023 | 203 | 387591 |
| 11/1/2018 | 61 | 211833 | 2/1/2021 | 1001 | 634578 | 5/1/2023 | 197 | 372595 |
| 12/1/2018 | 75 | 231077 | 3/1/2021 | 602 | 636611 | 6/1/2023 | 165 | 357262 |
| 1/1/2019 | 84 | 247048 | 4/1/2021 | 512 | 644706 | 7/1/2023 | 181 | 340671 |
| 2/1/2019 | 97 | 260750 | 5/1/2021 | 477 | 654532 | 8/1/2023 | 190 | 315175 |
| 3/1/2019 | 70 | 275500 | 6/1/2021 | 488 | 669422 | 9/1/2023 | 191 | 284909 |
| 4/1/2019 | 65 | 294571 | 7/1/2021 | 442 | 650654 | 10/1/2023 | 183 | 253365 |
| 5/1/2019 | 111 | 315796 | 8/1/2021 | 445 | 633474 | 11/1/2023 | 170 | 220501 |
| 6/1/2019 | 229 | 340147 | 9/1/2021 | 481 | 615328 | 12/1/2023 | 142 | 187869 |
| 7/1/2019 | 181 | 365951 | 10/1/2021 | 480 | 597309 | 1/1/2024 | 129 | 158416 |
| 8/1/2019 | 161 | 389629 | 11/1/2021 | 460 | 582057 | 2/1/2024 | 84 | 127248 |
| 9/1/2019 | 224 | 411682 | 12/1/2021 | 435 | 567977 | 3/1/2024 | 68 | 105243 |

| 10/1/2019 | 182 | 426396 | 1/1/2022 | 472 | 555753 | 4/1/2024 | 51 | 84075 |
|---|---|---|---|---|---|---|---|---|
| 11/1/2019 | 222 | 441027 | 2/1/2022 | 456 | 546166 | 5/1/2024 | 26 | 55394 |
| 12/1/2019 | 239 | 455225 | 3/1/2022 | 383 | 537980 | 6/1/2024 | 22 | 28900 |
| 1/1/2020 | 257 | 470692 | 4/1/2022 | 301 | 529346 | | | |
| 2/1/2020 | 204 | 480838 | 5/1/2022 | 295 | 520150 | | | |
| 3/1/2020 | 227 | 490628 | 6/1/2022 | 324 | 509888 | | | |
| 4/1/2020 | 722 | 500374 | 7/1/2022 | 344 | 498193 | | | |

## Appendix III: Algorithm for Selecting Bad and Good Loans

### Algorithm for Selecting Bad Loans

```
     if status=bad and      1 <= count_of_bads <=  500  and random_number<=1      then weight=1/1
else if status=bad and   501 <= count_of_bads <= 1000 and random_number<=0.95 then weight=1/0.95
else if status=bad and 1001 <= count_of_bads <= 2000 and random_number<=0.90 then weight=1/0.90
else if status=bad and 2001 <= count_of_bads <= 3000 and random_number<=0.85 then weight=1/0.85
else if status=bad and 3001 <= count_of_bads <= 4000 and random_number<=0.80 then weight=1/0.80
else if status=bad and 4001 <= count_of_bads <= 5000 and random_number<=0.75 then weight=1/0.75
else if status=bad and 5001 <= count_of_bads <= 6000 and random_number<=0.70 then weight=1/0.70
else if status=bad and 6001 <= count_of_bads              and random_number<=0.65 then weight=1/0.65
```

### Algorithm for Selecting Good Loans

```
     if status=good and          1 <= count_of_goods <= 100000 and random_number<=0.1               then weight=1/0.1
else if status=good and 100001 <= count_of_goods <= 200000 and random_number<=0.1               then weight=1/0.1
else if status=good and 200001 <= count_of_goods <= 300000 and random_number<=0.05             then weight=1/0.05
else if status=good and 300001 <= count_of_goods <= 400000 and random_number<=0.033333333 then weight=1/0.033333333
else if status=good and 400001 <= count_of_goods <= 500000 and random_number<=0.025           then weight=1/0.025
else if status=good and 500001 <= count_of_goods <= 600000 and random_number<=0.02             then weight=1/0.02
else if status=good and 600001 <= count_of_goods <= 700000 and random_number<=0.016666667   then weight=1/0.016666667
else if status=good and 700001 <= count_of_goods <= 800000 and random_number<=0.014285714   then weight=1/0.014285714
else if status=good and 800001 <= count_of_goods <= 900000 and random_number<=0.0125           then weight=1/0.0125
else if status=good and 900001 <= count_of_goods                  and random_number<=0.011111111    then weight=1/0.011111111
```

## Appendix IV: Bivariate Plots of Log-Odds of Bad Rate against Explanatory Variables

### Log-Odds of Bad Rate vs. Loan Attributes

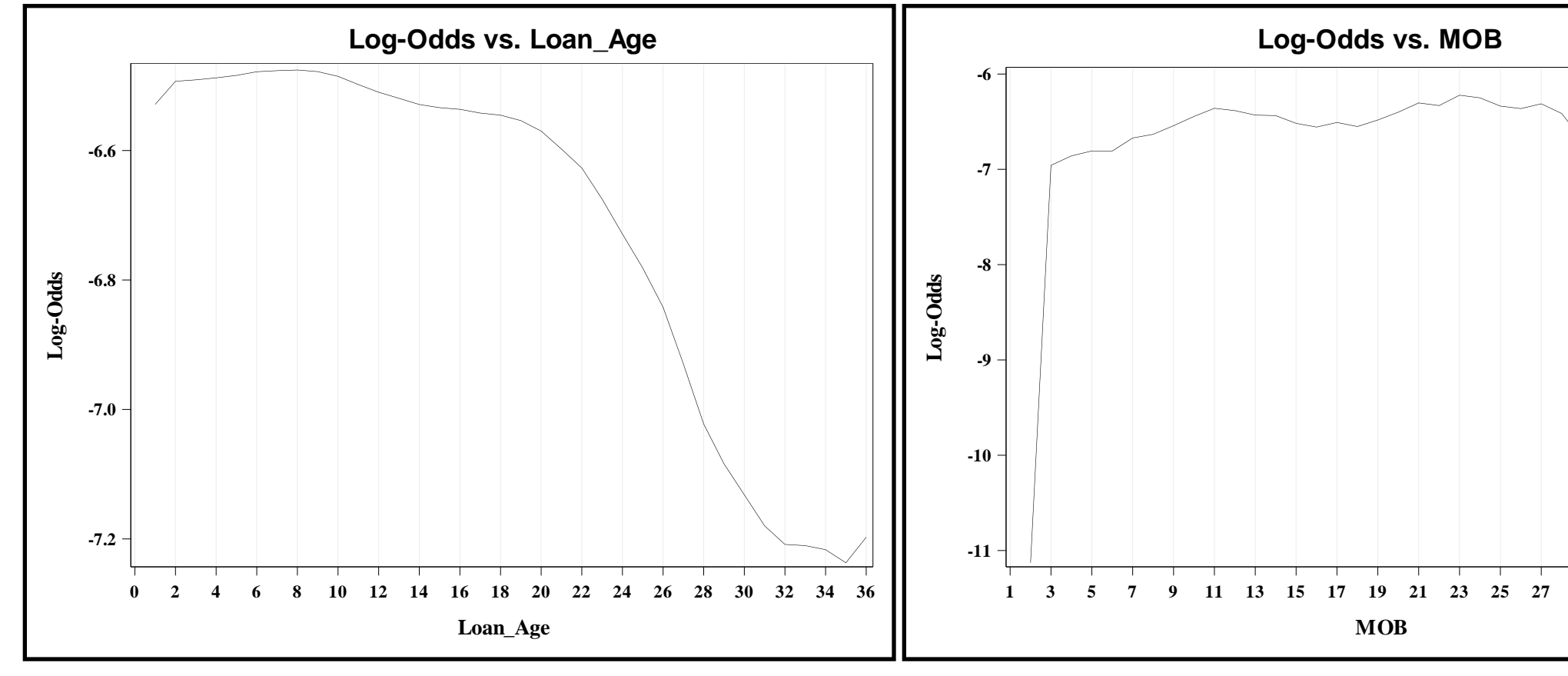

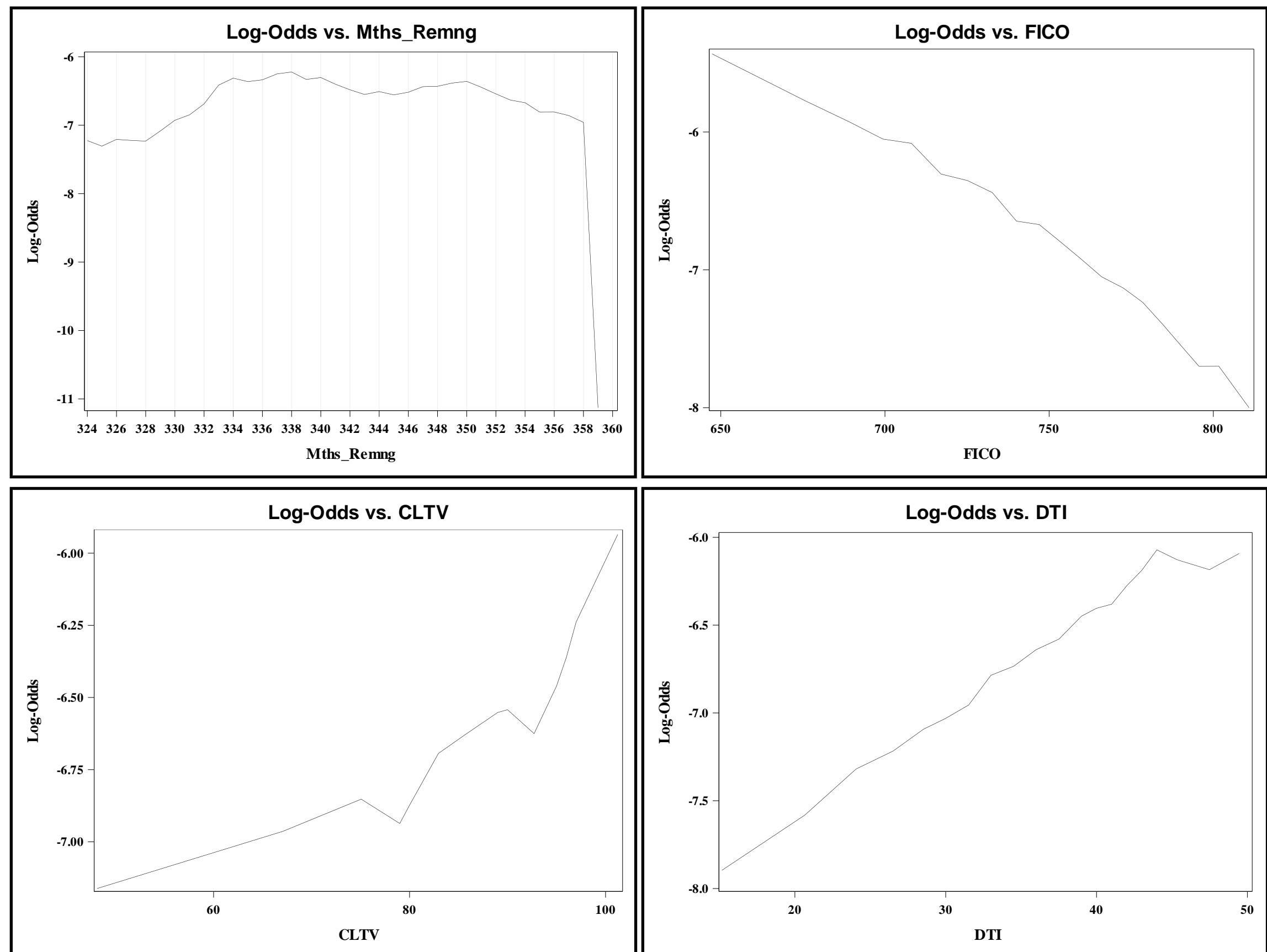

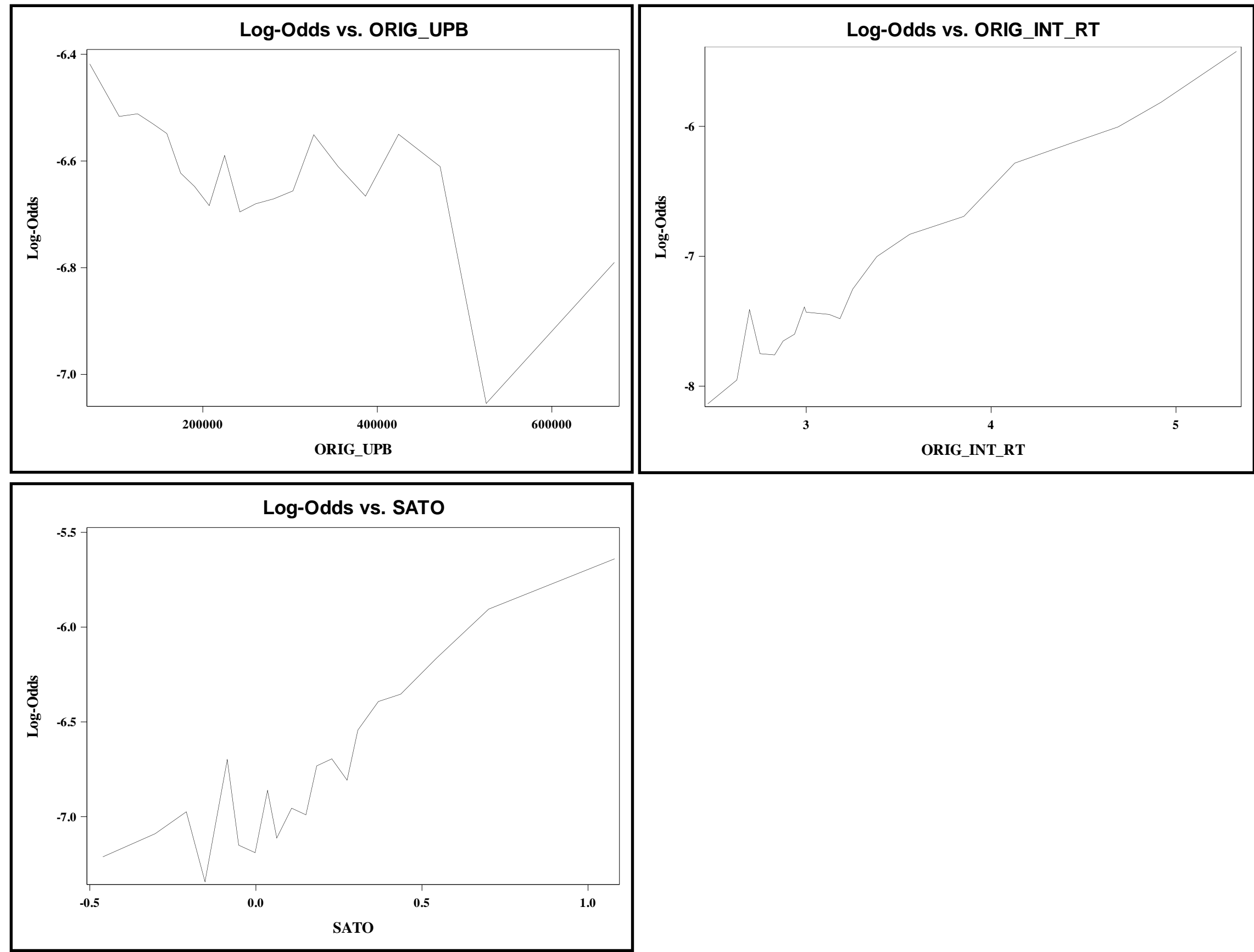

Log-odds of Bad Rate vs. Macroeconomic Variables

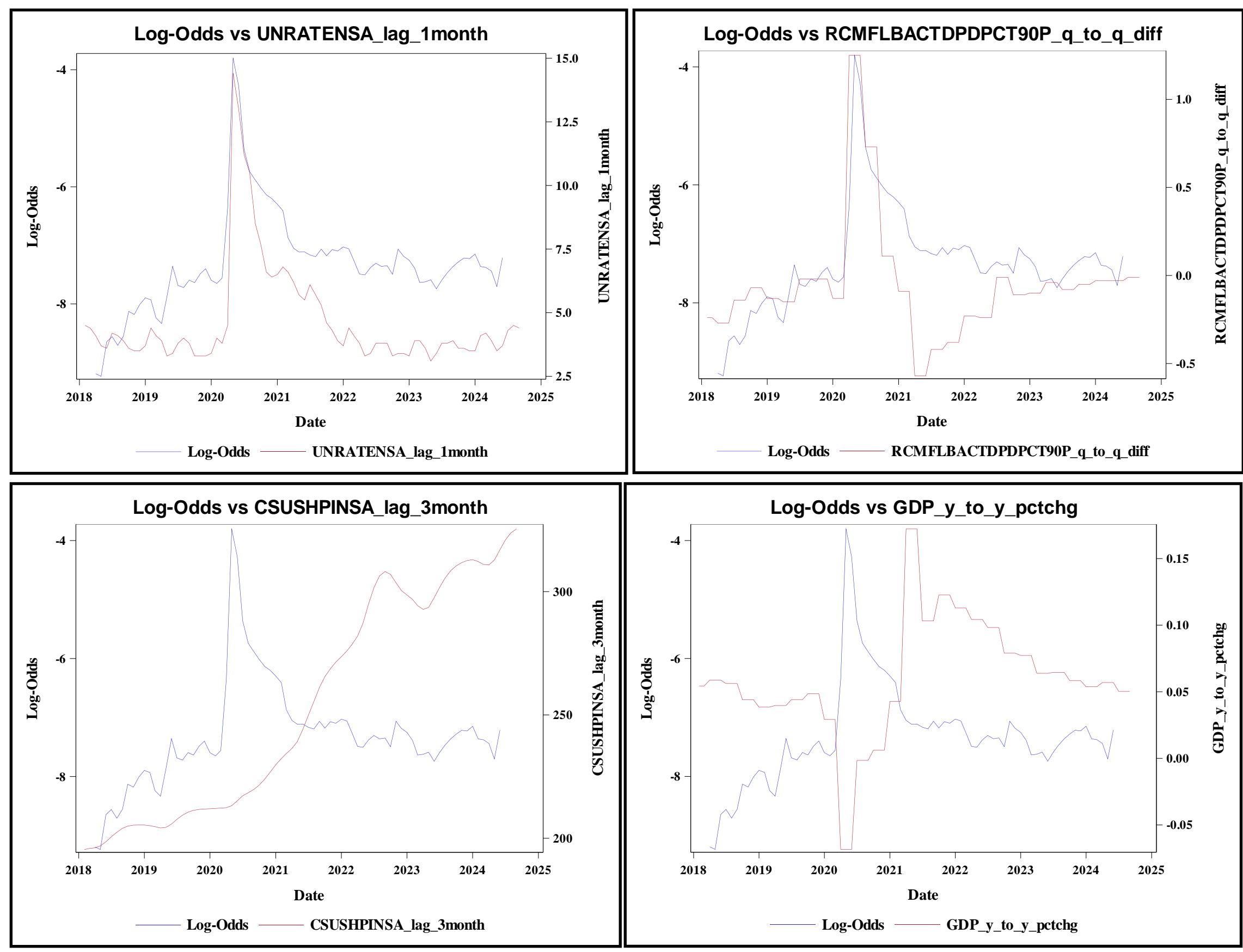

## Appendix V: Construction of The Interaction Term between FICO and Orig_UpB

To assess the potential interaction between FICO scores and original unpaid balances (Orig_UpB), FICO is partitioned into five categories and Orig_UpB into four. For each Orig_UpB group, the default rates are plotted across the FICO categories. The resulting visualization confirms the existence of an interaction effect between these two variables. Refer to the two figures in Section 4.5.

If the four curves are piecewise parallel or do not intersect, no interaction effect is present. To mitigate any observed interaction, specific constants can be determined; when the FICO and Orig_UpB variables are multiplied by these values, the resulting curves become piecewise parallel. The data utilized to generate the preceding plot are provided below.

| Orig_UpB_Group | FICO_Group1 | FICO_Group2 | FICO_Group3 | FICO_Group4 | FICO_Group5 |
|---|---|---|---|---|---|
| 1 | 0.002860102 | 0.001599081 | 0.001097379 | 0.000797963 | 0.00042826 |
| 2 | 0.002788828 | 0.001580301 | 0.001075895 | 0.000718272 | 0.00046151 |
| 3 | 0.0031314 | 0.001676672 | 0.001191523 | 0.000754367 | 0.000437466 |
| 4 | 0.002933905 | 0.001737331 | 0.001312948 | 0.000768765 | 0.000392404 |
| Average | 0.002928559 | 0.001648346 | 0.001169437 | 0.000759842 | 0.00042991 |

For each Orig_UpB group, the slope between FICO group 1 and FICO group 5, assuming a unit scale on the x-axis, is presented in the following table.

| Orig_UpB_Group | Slope between FICO group 1 and group 5 |
|---|---|
| 1 | 0.00060796 |
| 2 | 0.00058183 |
| 3 | 0.000673483 |
| 4 | 0.000635375 |
| Average | 0.000624662 |

To normalize each curve to match the average slope, the specific multipliers required for each group are provided in the following table.

| Orig_UpB_Group | Slope between FICO group 1 and group 5 | Multiplier |
|---|---|---|
| 1 | 0.00060796 | 1.027471902 |
| 2 | 0.00058183 | 1.073617097 |
| 3 | 0.000673483 | 0.927509341 |
| 4 | 0.000635375 | 0.983138827 |

By applying each multiplier to its corresponding slope, all slopes are normalized to the average value of 0.000624662. The following code is subsequently used to generate the interaction term between FICO and Orig_UpB.

```
if                Orig_UpB < 166000 then FICO_ Orig_UpB = FICO × 1.027471902
if 166000 <= Orig_UpB < 251000 then FICO _ Orig_UpB = FICO × 1.073617097
if 251000 <= Orig_UpB < 370000 then FICO _ Orig_UpB = FICO × 0.927509341
if 370000 <= Orig_UpB                then FICO _ Orig_UpB = FICO × 0.983138827
```

The results indicate a linear relationship between the interaction term and the default rate, as illustrated in the plot below.

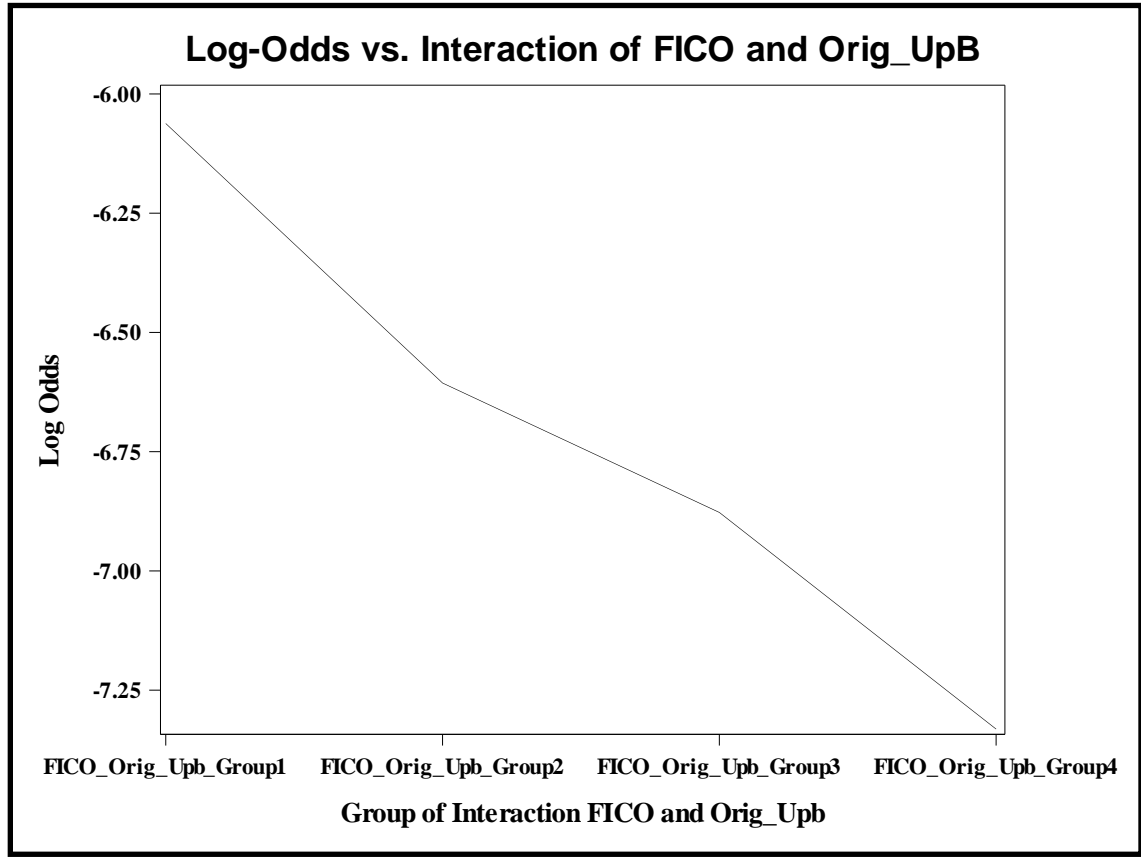

# Appendix VI: Results of GEE

# Data of Dropping Loan_Age_Basis

| Working Correlation Matrix Structure: Independent | | | | | | |
|---|---|---|---|---|---|---|
| | Empirical Standard Error Estimates | | | Model-Based Standard Error Estimates | | |
| Parameter | Estimate | Standard Error | p-value | Estimate | Standard Error | p-value |
| Intercept | -0.409 | 0.0639 | <.0001 | -0.409 | 0.0641 | <.0001 |
| mob_Basis1 | 0.0206 | 0.0016 | <.0001 | 0.0206 | 0.0015 | <.0001 |
| mob_Basis2 | 0.0299 | 0.0008 | <.0001 | 0.0299 | 0.0008 | <.0001 |
| mths_remng_Basis1 | -0.0645 | 0.0013 | <.0001 | -0.0645 | 0.0013 | <.0001 |
| mths_remng_Basis2 | -4.279 | 0.1902 | <.0001 | -4.279 | 0.1914 | <.0001 |
| cltv_Basis1 | 0.0197 | 0.0002 | <.0001 | 0.0197 | 0.0003 | <.0001 |
| cltv_Basis2 | -0.0046 | 0.0002 | <.0001 | -0.0046 | 0.0002 | <.0001 |
| fico | -0.0067 | 0 | <.0001 | -0.0067 | 0 | <.0001 |
| dti | 0.0296 | 0.0002 | <.0001 | 0.0296 | 0.0002 | <.0001 |
| sato | 0.1818 | 0.0065 | <.0001 | 0.1818 | 0.0066 | <.0001 |
| fico_ORIG_UPB | -0.0009 | 0 | <.0001 | -0.0009 | 0 | <.0001 |
| orig_int_rt | 0.1896 | 0.0055 | <.0001 | 0.1896 | 0.0056 | <.0001 |
| UNRATENSA_lag_1month | 0.3084 | 0.0011 | <.0001 | 0.3084 | 0.0011 | <.0001 |
| RCMFLBACTDPDPCT90P_q_to_q_pctchg | 0.6393 | 0.0155 | <.0001 | 0.6393 | 0.014 | <.0001 |
| GDP_y_to_y_pctchg | 4.9635 | 0.0632 | <.0001 | 4.9635 | 0.0624 | <.0001 |
| CSUSHPINSA_lag_3month | -0.0051 | 0.0001 | <.0001 | -0.0051 | 0.0001 | <.0001 |
| month2 | -0.3052 | 0.0066 | <.0001 | -0.3052 | 0.0066 | <.0001 |
| month3 | -0.5169 | 0.0071 | <.0001 | -0.5169 | 0.0071 | <.0001 |
| month4 | -0.5348 | 0.0075 | <.0001 | -0.5348 | 0.0076 | <.0001 |
| month5 | -0.675 | 0.0076 | <.0001 | -0.675 | 0.0072 | <.0001 |
| month6 | -0.5481 | 0.0074 | <.0001 | -0.5481 | 0.007 | <.0001 |
| month7 | -0.5859 | 0.0068 | <.0001 | -0.5859 | 0.0065 | <.0001 |
| month8 | -0.6005 | 0.0068 | <.0001 | -0.6005 | 0.0066 | <.0001 |
| month9 | -0.318 | 0.0063 | <.0001 | -0.318 | 0.0063 | <.0001 |
| month10 | -0.1726 | 0.0064 | <.0001 | -0.1726 | 0.0064 | <.0001 |
| month11 | 0.0323 | 0.0062 | <.0001 | 0.0323 | 0.0062 | <.0001 |

| Working Correlation Matrix Structure: Exchangeable | | | | | | |
|---|---|---|---|---|---|---|
| | Empirical Standard Error Estimates | | | Model-Based Standard Error Estimates | | |
| Parameter | Estimate | Standard Error | p-value | Estimate | Standard Error | p-value |
| Intercept | -0.355 | 0.0642 | <.0001 | -0.355 | 0.0638 | <.0001 |
| mob_Basis1 | 0.0206 | 0.0016 | <.0001 | 0.0206 | 0.0015 | <.0001 |
| mob_Basis2 | 0.0293 | 0.0008 | <.0001 | 0.0293 | 0.0008 | <.0001 |
| mths_remng_Basis1 | -0.064 | 0.0013 | <.0001 | -0.064 | 0.0013 | <.0001 |
| mths_remng_Basis2 | -4.2559 | 0.1853 | <.0001 | -4.2559 | 0.1888 | <.0001 |
| cltv_Basis1 | 0.0198 | 0.0002 | <.0001 | 0.0198 | 0.0003 | <.0001 |
| cltv_Basis2 | -0.0046 | 0.0002 | <.0001 | -0.0046 | 0.0002 | <.0001 |
| fico | -0.0068 | 0 | <.0001 | -0.0068 | 0 | <.0001 |
| dti | 0.0296 | 0.0002 | <.0001 | 0.0296 | 0.0002 | <.0001 |
| sato | 0.1844 | 0.0066 | <.0001 | 0.1844 | 0.0066 | <.0001 |
| fico_ORIG_UPB | -0.0009 | 0 | <.0001 | -0.0009 | 0 | <.0001 |
| orig_int_rt | 0.1848 | 0.0056 | <.0001 | 0.1848 | 0.0056 | <.0001 |
| UNRATENSA_lag_1month | 0.3093 | 0.0011 | <.0001 | 0.3093 | 0.0011 | <.0001 |
| RCMFLBACTDPDPCT90P_q_to_q_pctchg | 0.6501 | 0.0157 | <.0001 | 0.6501 | 0.0141 | <.0001 |
| GDP_y_to_y_pctchg | 5.0086 | 0.0641 | <.0001 | 5.0086 | 0.0628 | <.0001 |
| CSUSHPINSA_lag_3month | -0.0052 | 0.0001 | <.0001 | -0.0052 | 0.0001 | <.0001 |
| month2 | -0.3068 | 0.0067 | <.0001 | -0.3068 | 0.0066 | <.0001 |
| month3 | -0.5205 | 0.0073 | <.0001 | -0.5205 | 0.0072 | <.0001 |
| month4 | -0.54 | 0.0076 | <.0001 | -0.54 | 0.0077 | <.0001 |
| month5 | -0.6819 | 0.0077 | <.0001 | -0.6819 | 0.0073 | <.0001 |
| month6 | -0.5547 | 0.0075 | <.0001 | -0.5547 | 0.0071 | <.0001 |
| month7 | -0.5897 | 0.0069 | <.0001 | -0.5897 | 0.0066 | <.0001 |
| month8 | -0.6045 | 0.0069 | <.0001 | -0.6045 | 0.0066 | <.0001 |
| month9 | -0.3202 | 0.0064 | <.0001 | -0.3202 | 0.0064 | <.0001 |
| month10 | -0.1736 | 0.0065 | <.0001 | -0.1736 | 0.0065 | <.0001 |
| month11 | 0.0327 | 0.0063 | <.0001 | 0.0327 | 0.0062 | <.0001 |

| Working Correlation Matrix Structure: AR(1) | | | | | | |
|---|---|---|---|---|---|---|
| | Empirical Standard Error Estimates | | | Model-Based Standard Error Estimates | | |
| Parameter | Estimate | Standard Error | p-value | Estimate | Standard Error | p-value |

| Intercept | -0.4113 | 0.0639 | <.0001 | -0.4113 | 0.064 | <.0001 |
|---|---|---|---|---|---|---|
| mob_Basis1 | 0.0207 | 0.0016 | <.0001 | 0.0207 | 0.0015 | <.0001 |
| mob_Basis2 | 0.0299 | 0.0008 | <.0001 | 0.0299 | 0.0008 | <.0001 |
| mths_remng_Basis1 | -0.0645 | 0.0013 | <.0001 | -0.0645 | 0.0013 | <.0001 |
| mths_remng_Basis2 | -4.2757 | 0.1897 | <.0001 | -4.2757 | 0.1911 | <.0001 |
| cltv_Basis1 | 0.0197 | 0.0002 | <.0001 | 0.0197 | 0.0003 | <.0001 |
| cltv_Basis2 | -0.0046 | 0.0002 | <.0001 | -0.0046 | 0.0002 | <.0001 |
| fico | -0.0067 | 0 | <.0001 | -0.0067 | 0 | <.0001 |
| dti | 0.0296 | 0.0002 | <.0001 | 0.0296 | 0.0002 | <.0001 |
| sato | 0.1817 | 0.0065 | <.0001 | 0.1817 | 0.0066 | <.0001 |
| fico_ORIG_UPB | -0.0009 | 0 | <.0001 | -0.0009 | 0 | <.0001 |
| orig_int_rt | 0.1896 | 0.0055 | <.0001 | 0.1896 | 0.0056 | <.0001 |
| UNRATENSA_lag_1month | 0.3084 | 0.0011 | <.0001 | 0.3084 | 0.0011 | <.0001 |
| RCMFLBACTDPDPCT90P_q_to_q_pctchg | 0.6392 | 0.0155 | <.0001 | 0.6392 | 0.014 | <.0001 |
| GDP_y_to_y_pctchg | 4.9639 | 0.0632 | <.0001 | 4.9639 | 0.0624 | <.0001 |
| CSUSHPINSA_lag_3month | -0.0051 | 0.0001 | <.0001 | -0.0051 | 0.0001 | <.0001 |
| month2 | -0.3053 | 0.0066 | <.0001 | -0.3053 | 0.0066 | <.0001 |
| month3 | -0.517 | 0.0071 | <.0001 | -0.517 | 0.0071 | <.0001 |
| month4 | -0.5346 | 0.0075 | <.0001 | -0.5346 | 0.0076 | <.0001 |
| month5 | -0.6745 | 0.0076 | <.0001 | -0.6745 | 0.0072 | <.0001 |
| month6 | -0.5484 | 0.0074 | <.0001 | -0.5484 | 0.007 | <.0001 |
| month7 | -0.5862 | 0.0068 | <.0001 | -0.5862 | 0.0065 | <.0001 |
| month8 | -0.6006 | 0.0068 | <.0001 | -0.6006 | 0.0066 | <.0001 |
| month9 | -0.318 | 0.0063 | <.0001 | -0.318 | 0.0063 | <.0001 |
| month10 | -0.1726 | 0.0064 | <.0001 | -0.1726 | 0.0064 | <.0001 |
| month11 | 0.0324 | 0.0062 | <.0001 | 0.0324 | 0.0062 | <.0001 |

# Data of Adding Additional Records

| **Working Correlation Matrix Structure: Independent** | | | | | | |
|---|---|---|---|---|---|---|
| | Empirical Standard Error Estimates | | | Model-Based Standard Error Estimates | | |
| Parameter | Estimate | Standard Error | p-value | Estimate | Standard Error | p-value |
| Intercept | -0.3887 | 0.0647 | <.0001 | -0.3887 | 0.0642 | <.0001 |
| Loan_Age_Basis | -9.3194 | 0.9982 | <.0001 | -9.3194 | 0.5605 | <.0001 |
| mob_Basis1 | 0.0328 | 0.0016 | <.0001 | 0.0328 | 0.0015 | <.0001 |
| mob_Basis2 | 0.0281 | 0.0008 | <.0001 | 0.0281 | 0.0008 | <.0001 |
| mths_remng_Basis1 | -0.0685 | 0.0013 | <.0001 | -0.0685 | 0.0013 | <.0001 |
| mths_remng_Basis2 | -3.746 | 0.1901 | <.0001 | -3.746 | 0.1909 | <.0001 |
| cltv_Basis1 | 0.0198 | 0.0003 | <.0001 | 0.0198 | 0.0003 | <.0001 |
| cltv_Basis2 | -0.0046 | 0.0002 | <.0001 | -0.0046 | 0.0002 | <.0001 |
| fico | -0.0068 | 0 | <.0001 | -0.0068 | 0 | <.0001 |
| dti | 0.0296 | 0.0002 | <.0001 | 0.0296 | 0.0002 | <.0001 |
| sato | 0.1759 | 0.0066 | <.0001 | 0.1759 | 0.0066 | <.0001 |
| fico_ORIG_UPB | -0.0009 | 0 | <.0001 | -0.0009 | 0 | <.0001 |
| orig_int_rt | 0.1938 | 0.0056 | <.0001 | 0.1938 | 0.0056 | <.0001 |
| UNRATENSA_lag_1month | 0.3088 | 0.0011 | <.0001 | 0.3088 | 0.0011 | <.0001 |
| RCMFLBACTDPDPCT90P_q_to_q_pctchg | 0.6348 | 0.0155 | <.0001 | 0.6348 | 0.014 | <.0001 |
| GDP_y_to_y_pctchg | 4.9415 | 0.0632 | <.0001 | 4.9415 | 0.0624 | <.0001 |
| CSUSHPINSA_lag_3month | -0.005 | 0.0001 | <.0001 | -0.005 | 0.0001 | <.0001 |
| month2 | -0.3049 | 0.0066 | <.0001 | -0.3049 | 0.0066 | <.0001 |
| month3 | -0.516 | 0.0071 | <.0001 | -0.516 | 0.0071 | <.0001 |
| month4 | -0.532 | 0.0075 | <.0001 | -0.532 | 0.0076 | <.0001 |
| month5 | -0.6734 | 0.0076 | <.0001 | -0.6734 | 0.0072 | <.0001 |
| month6 | -0.5465 | 0.0074 | <.0001 | -0.5465 | 0.0071 | <.0001 |
| month7 | -0.5862 | 0.0068 | <.0001 | -0.5862 | 0.0066 | <.0001 |
| month8 | -0.601 | 0.0068 | <.0001 | -0.601 | 0.0066 | <.0001 |
| month9 | -0.3188 | 0.0063 | <.0001 | -0.3188 | 0.0063 | <.0001 |
| month10 | -0.1735 | 0.0064 | <.0001 | -0.1735 | 0.0064 | <.0001 |
| month11 | 0.0317 | 0.0062 | <.0001 | 0.0317 | 0.0062 | <.0001 |

| **Working Correlation Matrix Structure: Exchangeable** | | | | | | |
|---|---|---|---|---|---|---|
| | Empirical Standard Error Estimates | | | Model-Based Standard Error Estimates | | |
| Parameter | Estimate | Standard Error | p-value | Estimate | Standard Error | p-value |
| Intercept | -0.3923 | 0.0647 | <.0001 | -0.3923 | 0.0642 | <.0001 |
| Loan_Age_Basis | -9.3369 | 1.0036 | <.0001 | -9.3369 | 0.5652 | <.0001 |
| mob_Basis1 | 0.0328 | 0.0016 | <.0001 | 0.0328 | 0.0015 | <.0001 |
| mob_Basis2 | 0.0281 | 0.0008 | <.0001 | 0.0281 | 0.0008 | <.0001 |
| mths_remng_Basis1 | -0.0685 | 0.0013 | <.0001 | -0.0685 | 0.0013 | <.0001 |
| mths_remng_Basis2 | -3.7461 | 0.1901 | <.0001 | -3.7461 | 0.1909 | <.0001 |
| cltv_Basis1 | 0.0198 | 0.0003 | <.0001 | 0.0198 | 0.0003 | <.0001 |
| cltv_Basis2 | -0.0046 | 0.0002 | <.0001 | -0.0046 | 0.0002 | <.0001 |
| fico | -0.0068 | 0 | <.0001 | -0.0068 | 0 | <.0001 |
| dti | 0.0296 | 0.0002 | <.0001 | 0.0296 | 0.0002 | <.0001 |
| sato | 0.1757 | 0.0066 | <.0001 | 0.1757 | 0.0066 | <.0001 |
| fico_ORIG_UPB | -0.0009 | 0 | <.0001 | -0.0009 | 0 | <.0001 |
| orig_int_rt | 0.1941 | 0.0056 | <.0001 | 0.1941 | 0.0056 | <.0001 |
| UNRATENSA_lag_1month | 0.3087 | 0.0011 | <.0001 | 0.3087 | 0.0011 | <.0001 |
| RCMFLBACTDPDPCT90P_q_to_q_pctchg | 0.634 | 0.0155 | <.0001 | 0.634 | 0.014 | <.0001 |
| GDP_y_to_y_pctchg | 4.9381 | 0.0632 | <.0001 | 4.9381 | 0.0624 | <.0001 |
| CSUSHPINSA_lag_3month | -0.005 | 0.0001 | <.0001 | -0.005 | 0.0001 | <.0001 |
| month2 | -0.3048 | 0.0066 | <.0001 | -0.3048 | 0.0066 | <.0001 |
| month3 | -0.5157 | 0.0071 | <.0001 | -0.5157 | 0.0071 | <.0001 |
| month4 | -0.5316 | 0.0075 | <.0001 | -0.5316 | 0.0076 | <.0001 |
| month5 | -0.6729 | 0.0076 | <.0001 | -0.6729 | 0.0072 | <.0001 |
| month6 | -0.546 | 0.0074 | <.0001 | -0.546 | 0.007 | <.0001 |
| month7 | -0.5859 | 0.0068 | <.0001 | -0.5859 | 0.0065 | <.0001 |
| month8 | -0.6007 | 0.0068 | <.0001 | -0.6007 | 0.0066 | <.0001 |
| month9 | -0.3186 | 0.0063 | <.0001 | -0.3186 | 0.0063 | <.0001 |
| month10 | -0.1734 | 0.0064 | <.0001 | -0.1734 | 0.0064 | <.0001 |
| month11 | 0.0317 | 0.0062 | <.0001 | 0.0317 | 0.0062 | <.0001 |

| Working Correlation Matrix Structure: AR(1) | | | | | | |
|---|---|---|---|---|---|---|
| | Empirical Standard Error Estimates | | | Model-Based Standard Error Estimates | | |
| Parameter | Estimate | Standard Error | p-value | Estimate | Standard Error | p-value |
| Intercept | -0.3874 | 0.0647 | <.0001 | -0.3874 | 0.0642 | <.0001 |
| Loan_Age_Basis | -9.3203 | 0.9976 | <.0001 | -9.3203 | 0.5606 | <.0001 |
| mob_Basis1 | 0.0328 | 0.0016 | <.0001 | 0.0328 | 0.0015 | <.0001 |
| mob_Basis2 | 0.0281 | 0.0008 | <.0001 | 0.0281 | 0.0008 | <.0001 |
| mths_remng_Basis1 | -0.0685 | 0.0013 | <.0001 | -0.0685 | 0.0013 | <.0001 |
| mths_remng_Basis2 | -3.7468 | 0.1902 | <.0001 | -3.7468 | 0.1909 | <.0001 |
| cltv_Basis1 | 0.0198 | 0.0003 | <.0001 | 0.0198 | 0.0003 | <.0001 |
| cltv_Basis2 | -0.0046 | 0.0002 | <.0001 | -0.0046 | 0.0002 | <.0001 |
| fico | -0.0068 | 0 | <.0001 | -0.0068 | 0 | <.0001 |
| dti | 0.0296 | 0.0002 | <.0001 | 0.0296 | 0.0002 | <.0001 |
| sato | 0.176 | 0.0066 | <.0001 | 0.176 | 0.0066 | <.0001 |
| fico_ORIG_UPB | -0.0009 | 0 | <.0001 | -0.0009 | 0 | <.0001 |
| orig_int_rt | 0.1937 | 0.0056 | <.0001 | 0.1937 | 0.0056 | <.0001 |
| UNRATENSA_lag_1month | 0.3088 | 0.0011 | <.0001 | 0.3088 | 0.0011 | <.0001 |
| RCMFLBACTDPDPCT90P_q_to_q_pctchg | 0.6349 | 0.0155 | <.0001 | 0.6349 | 0.014 | <.0001 |
| GDP_y_to_y_pctchg | 4.9413 | 0.0632 | <.0001 | 4.9413 | 0.0624 | <.0001 |
| CSUSHPINSA_lag_3month | -0.005 | 0.0001 | <.0001 | -0.005 | 0.0001 | <.0001 |
| month2 | -0.3049 | 0.0066 | <.0001 | -0.3049 | 0.0066 | <.0001 |
| month3 | -0.5159 | 0.0071 | <.0001 | -0.5159 | 0.0071 | <.0001 |
| month4 | -0.5321 | 0.0075 | <.0001 | -0.5321 | 0.0076 | <.0001 |
| month5 | -0.6737 | 0.0076 | <.0001 | -0.6737 | 0.0072 | <.0001 |
| month6 | -0.5464 | 0.0074 | <.0001 | -0.5464 | 0.0071 | <.0001 |
| month7 | -0.586 | 0.0068 | <.0001 | -0.586 | 0.0065 | <.0001 |
| month8 | -0.6009 | 0.0068 | <.0001 | -0.6009 | 0.0066 | <.0001 |
| month9 | -0.3188 | 0.0063 | <.0001 | -0.3188 | 0.0063 | <.0001 |
| month10 | -0.1735 | 0.0064 | <.0001 | -0.1735 | 0.0064 | <.0001 |
| month11 | 0.0317 | 0.0062 | <.0001 | 0.0317 | 0.0062 | <.0001 |

## Appendix VII: Back test plots of Selective Snapshots (In-Sample and Holdout)

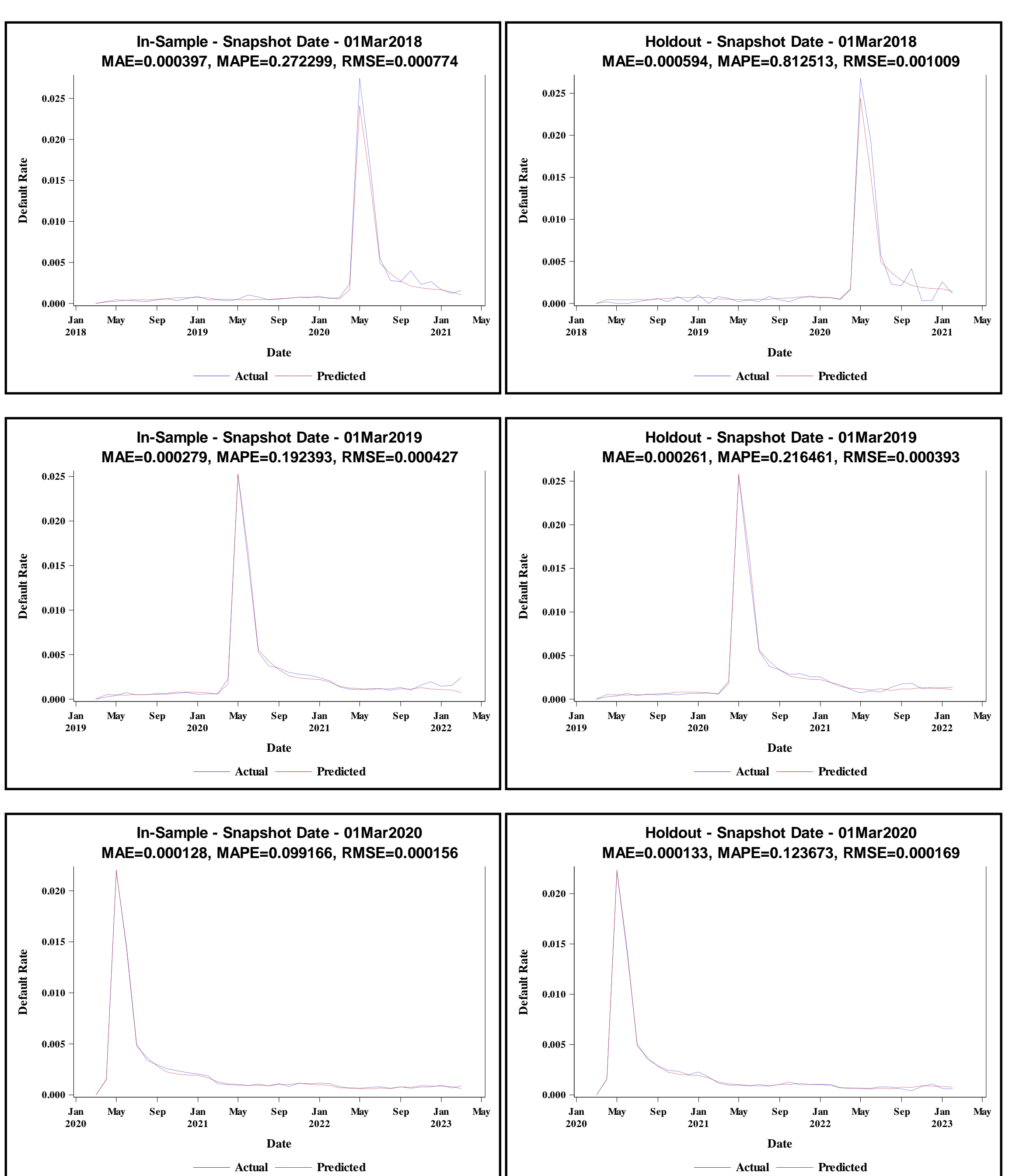

## Appendix VIII: AUC, Rank Order Table and Back Test Plot of Scorecard Samples of In-Sample and Out-of-Time Data

| AUC of In-Sample Scorecard Sample | AUC of Out-of-Time Scorecard Sample |
| --- | --- |

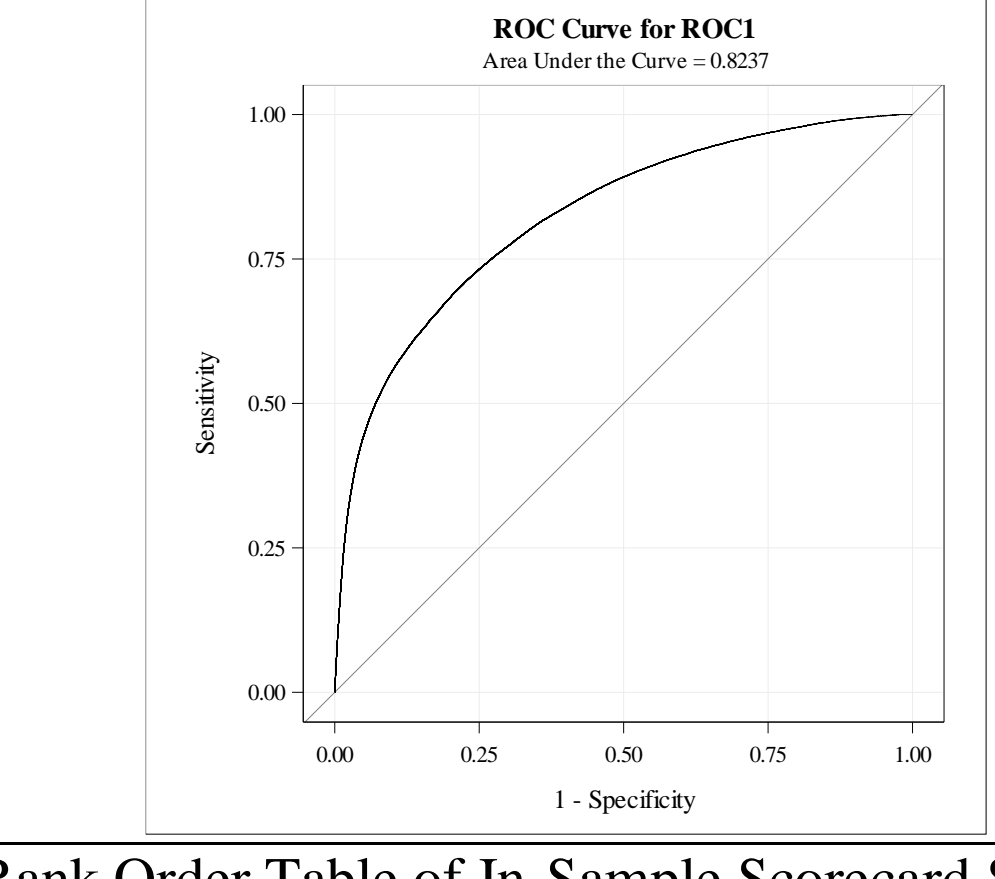

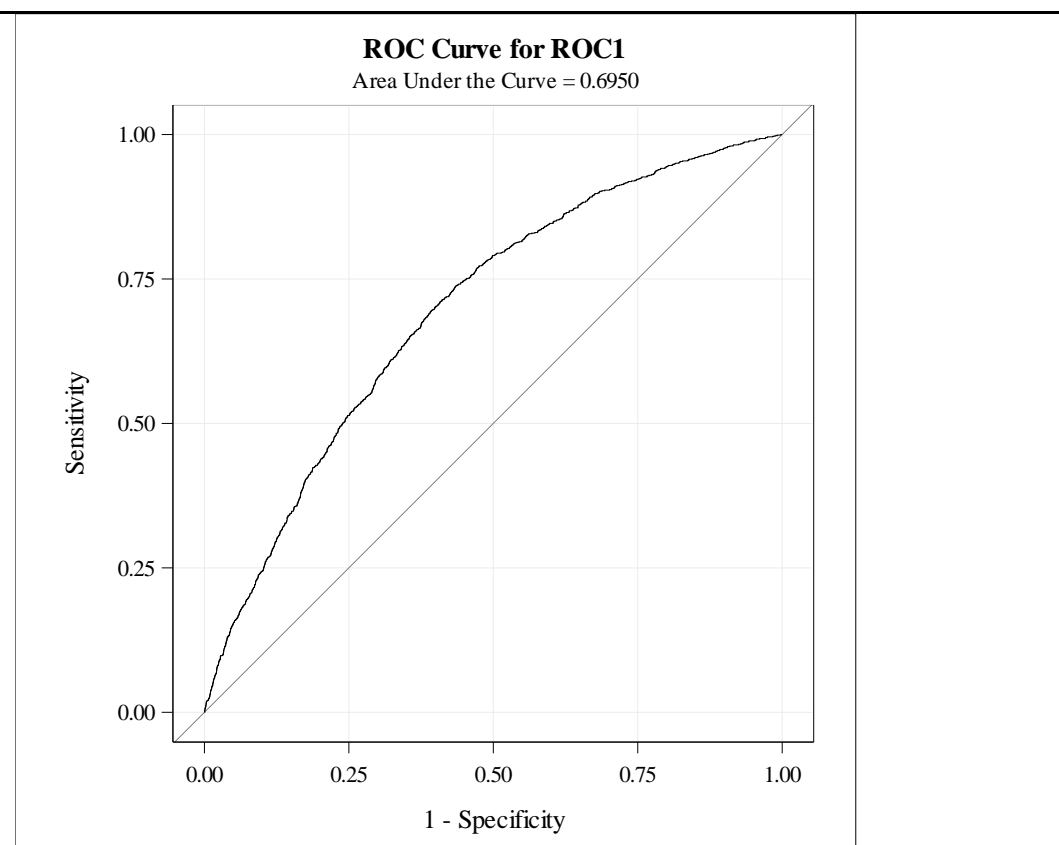

Rank Order Table of In-Sample Scorecard Sample

| Decile | Default Pred | Default Actual |
|---|---|---|
| 1 | 0.0006 | 0.0004 |
| 2 | 0.0015 | 0.0011 |
| 3 | 0.0021 | 0.0015 |
| 4 | 0.0028 | 0.0021 |
| 5 | 0.0036 | 0.0030 |
| 6 | 0.0046 | 0.0042 |
| 7 | 0.0060 | 0.0058 |
| 8 | 0.0081 | 0.0080 |
| 9 | 0.0128 | 0.0119 |
| 10 | 0.0619 | 0.0594 |

Rank Order Table of Out-of-Time Scorecard Sample

| Decile | Default Pred | Default Actual |
|---|---|---|
| 1 | 0.0008 | 0.0009 |
| 2 | 0.0014 | 0.0012 |
| 3 | 0.0018 | 0.0015 |
| 4 | 0.0023 | 0.0021 |
| 5 | 0.0028 | 0.0021 |
| 6 | 0.0034 | 0.0033 |
| 7 | 0.0042 | 0.0045 |
| 8 | 0.0053 | 0.0053 |
| 9 | 0.0069 | 0.0069 |
| 10 | 0.0109 | 0.0090 |

Backtest Plot of Scorecard Sample

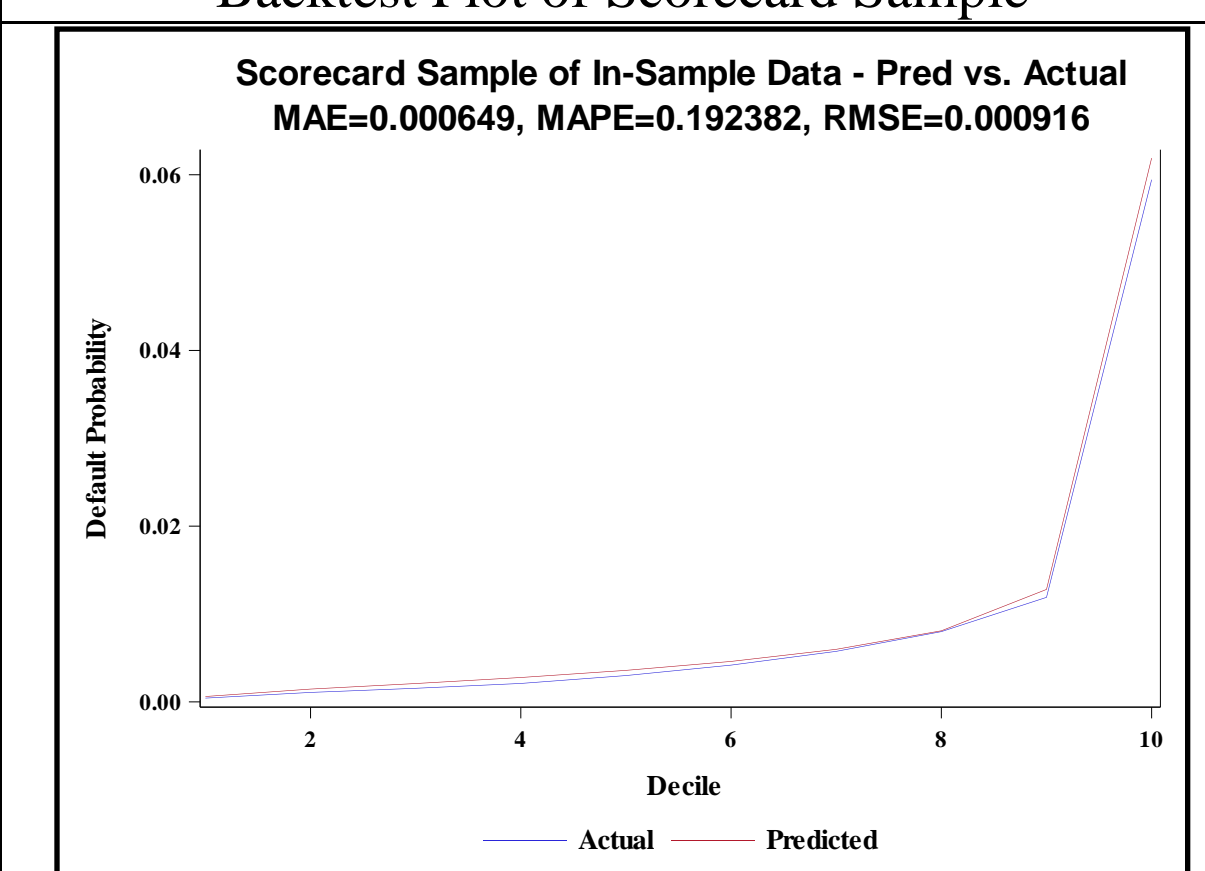

Backtest Plot of Out-of-Time Scorecard Sample

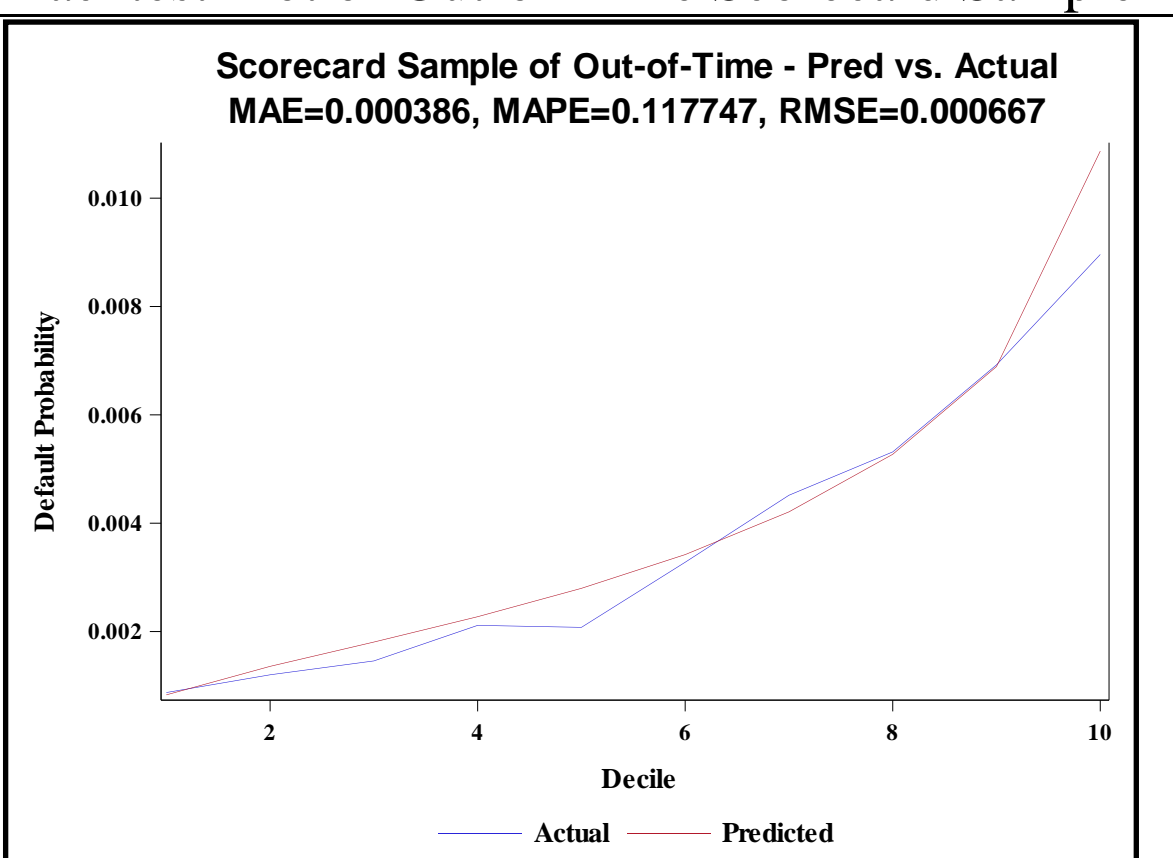